# Minimum Sample Size for Developing a Multivariable Prediction Model using Multinomial Logistic Regression

Alexander Pate, PhD [1], Richard D Riley, PhD [2], Gary S Collins, PhD [3,4], Maarten van Smeden, PhD [5,6], Ben Van Calster, PhD [7,8,9], Joie Ensor, PhD [2], Glen P Martin, PhD [1]

1. Division of Informatics, Imaging and Data Science, Faculty of Biology, Medicine and Health, University of Manchester, Manchester Academic Health Science Centre, Manchester, UK

2. Centre for Prognosis Research, School of Medicine, Keele University, Staffordshire, UK

3. Centre for Statistics in Medicine, Nuffield Department of Orthopaedics, Rheumatology and Musculoskeletal Sciences, University of Oxford, Oxford, UK

4. NIHR Oxford Biomedical Research Centre, John Radcliffe Hospital, Oxford, UK.

5. Julius Center for Health Sciences, University Medical Center Utrecht, Utrecht University, Utrecht, Netherlands

6. Department of Clinical Epidemiology, Leiden University Medical Center Leiden, Netherlands

7. Department of Development and Regeneration, KU Leuven, Leuven, Belgium

8. Department of Biomedical Data Sciences, Leiden University Medical Centre, the Netherlands

9. EPI-center, KU Leuven, Leuven, Belgium

**Running Title:** Minimum Sample Size for Multinomial Logistic Regression

**Corresponding Author:**

Dr Alexander Pate

Jean McFarlane, University of Manchester, Manchester, M13 9GB, United Kingdom

Email: alexander.pate@manchester.ac.uk

Word count (excluding references): 6,122

// Abstract

Multinomial logistic regression models allow one to predict the risk of a categorical outcome with > 2 categories. When developing such a model, researchers should ensure the number of participants ($n$) is appropriate relative to the number of events ($E_k$) and the number of predictor parameters ($p_k$) for each category $k$. We propose three criteria to determine the minimum $n$ required in light of existing criteria developed for binary outcomes. The first criteria aims to minimise the model overfitting. The second aims to minimise the difference between the observed and adjusted $R^2$ Nagelkerke. The third criterion aims to ensure the overall risk is estimated precisely. For criterion (i), we show the sample size must be based on the anticipated Cox-snell $R^2$ of distinct "one-to-one" logistic regression models corresponding to the sub-models of the multinomial logistic regression, rather than on the overall Cox-snell $R^2$ of the multinomial logistic regression. We tested the performance of the proposed criteria (i) through a simulation study, and found that it resulted in the desired level of overfitting. Criterion (ii) and (iii) are natural extensions from previously proposed criteria for binary outcomes. We illustrate how to implement the sample size criteria through a worked example considering the development of a multinomial risk prediction model for tumour type when presented with an ovarian mass. Code is provided for the simulation and worked example. We will embed our proposed criteria within the pmsampsize R library and Stata modules.

## Keywords

Clinical Prediction Models, Sample Size, Multinomial Logistic Regression, Shrinkage

## 1. Introduction

Clinical prediction models (CPMs) are developed to predict expected health outcomes, such as an individual's probability that a specific disease or condition is present (diagnostic models) or that a specific event will occur in the future (prognostic models).[1,2] Logistic regression is typically used for developing CPMs to predict a single binary outcome. Often though, healthcare outcomes have multiple levels (multi-category/ polytomous outcomes), such as cancer grade or Likert scales. Then, the natural extension is to use multinomial logistic regression to develop the CPM. Multinomial models have been used to develop CPMs across a range of clinical settings,[3–8] and it has been argued they should be used to develop prediction models more often.[9] It has also been shown that multinomial regression is preferred over multiple binary logistic regression when predicting multiple correlated binary outcomes to estimate their joint probability.[10]

An important design aspect when developing any prediction model is ensuring the sample size of the development dataset is appropriate to minimise overfitting and ensure sufficiently precise predictions. Sample size guidance for developing prediction models with continuous, binary and time-to-event outcomes have recently been developed.[11–15] However, there is a paucity of guidance for multinomial prediction models. Work by de Jong et al.,[16] highlighted the importance of considering the number of events per predictor for each outcome category when choosing the sample size, and showed that multinomial logistic regression models were susceptible to overfitting when fit in development data of small-to-medium sample size. However, there is no evidence to support events per predictor rules-of-thumb for calculating required sample size,[13,17] and more tailored guidance is required.

Therefore, the aim of this study was to extend the existing sample size criteria by Riley et al.,[11,12] to cater for multinomial risk prediction models. The remainder of this paper is structured as follows: Section 2 briefly reviews the minimum sample size criterion outlined by Riley et al. for binary CPMs,[12] and Section 3 uses these as the foundation for our proposed sample size criteria for developing a multinomial logistic regression model. A detailed description of the simulation used to verify one of the proposed sample size criteria is given in Appendix S1 (section 1). Sections 4 and 5 illustrate how to use the proposed criteria in practice, through discussion of how to use information from published models as the basis for the calculations (section 4) and through a worked example (section 5). Finally, in Section 6 we summarise the findings.

## 2. Existing sample size proposal for developing prediction models using binary logistic regression

We use the sample size criteria proposed by Riley et al.[12] as the basis for our extensions into multinomial logistic regression. In this section we introduce the notation required for our proposals, but refer readers to previous literature[11,12] for a full discussion.

Given a binary outcome, $Y_i$ $(i = 1, \ldots, N)$, which takes the value 1 if observation $i$ has the outcome and is 0 otherwise. CPMs for such outcomes aim to estimate the probability of $Y_i = 1$ conditional on a set of $Q$ (candidate) predictor parameters, denoted as $X_{qi}$ for $q = 1, \ldots, Q$, collectively in the vector $X_i = \left(X_{1i}, \ldots X_{Qi}\right)^T$. This can be modelled using logistic regression to estimate $P(Y_i = 1|X_i)$, as

$$\log\left(\frac{P(Y_i = 1|X_i)}{1 - P(Y_i = 1|X_i)}\right) = \beta_0 + \beta_1 X_{1i} + \cdots + \beta_Q X_{Qi}\,, \qquad (1)$$

where $\beta_1, \dots, \beta_Q$ are a set of predictor coefficients (conditional log odds ratios), which are estimated through maximum likelihood estimation to give estimates $\hat{\beta}_1, \dots, \hat{\beta}_Q$.

The Riley et al. sample size criteria for developing a binary CPM based on equation (1) have three components: (i) targeting a suitable shrinkage factor, (ii) targeting a small absolute difference in the apparent and adjusted Nagelkerke's $R^2$ ($R^2_{\text{Nagelkerke}}$)[18], and (iii) targeting a precise estimate of overall risk (model intercepts). We now overview these components in turn.

## 2.1. Overview of criterion (i): sample size required to minimise overfitting of predictor effects

The first sample size criterion of Riley et al.[12] assesses overfitting on the multiplicative scale by considering shrinkage of predictor effects. This is when regression coefficients are shrunk towards zero to help mitigate against risk of overfitting. Criterion (i) is based on a global shrinkage factor (S) that is applied to all predictor effects. Specifically, one multiplies $\hat{\beta}_1, \dots, \hat{\beta}_Q$ of equation (1) by *S*, giving,

$$\log\left(\frac{P(Y_i = 1|X_i)}{1 - P(Y_i = 1|X_i)}\right) = \alpha^* + S(\hat{\beta}_1 X_{1i} + \cdots + \beta_Q X_{Qi})\,, \tag{2}$$

where $\alpha^*$is the revised intercept to ensure the mean predicted risk matches the mean observed risk.[19] For the sample size criteria,[12] the van Houwelingen and Le Cessie's heuristic shrinkage factor ($S_{VH}$)[20] is used to estimate $S$:

$$S_{VH} = 1 - \frac{Q}{LR}\,, \tag{3}$$

where $Q$ is the number of predictors parameters, and $LR = -2(lnL_{null} - lnL_{model})$ is the likelihood ratio statistic.

Criterion (i) of Riley et al.,[12] calculates a sample size $n$ to target the shrinkage ($S_{VH}$) to be above a pre-specified threshold (commonly taken as 0.9 or above, to target an overfitting of 10% or less, which leads to greater model stability[21,22]). For binary logistic regression, the required sample size to target a shrinkage factor $S_{VH}$, is calculated using equation (4):

$$n = \frac{Q}{(S_{VH} - 1)\log\left(1 - \frac{R^2_{CS_adj}}{S_{VH}}\right)}\,, \tag{4}$$

where $Q$ is the number of candidate predictor parameters, and $R^2_{CS_adj}$, is an optimism adjusted estimate of the Cox-Snell[23] $R^2_{CS}$.

## 2.2. Overview of criterion (ii): ensuring small absolute difference in the apparent and adjusted $R^2_{\text{Nagelkerke}}$

The second sample size criterion of Riley et al.[12] is defined to ensure a small difference ($\delta$) between the apparent and adjusted Nagelkerke $R^2$. It requires pre-specifying a value for $\delta$ that one would tolerate, with small values preferred to improve model stability.[21,22] For any generalised linear model, Nagelkerke $R^2$ is expressed as:

$$R^2_{\text{Nagelkerke}} = \frac{R^2_{CS}}{\max\,(R^2_{CS})}\,, \tag{5}$$

where $R^2_{CS}$ could be the apparent or optimism adjusted estimate of $R^2_{CS}$. The maximum value of $R^2_{CS}$ is calculated as $\max(R^2_{CS}) = \max\left(R^2_{CS_app}\right) = \max\left(R^2_{CS_adj}\right) = 1 - \exp\left(\frac{2\, ln\, L_{null}}{n}\right)$, where $ln\, L_{null}$ is the log-likehood of the intercept only model. It then follows that

$$R^2_{\text{Nagelkerke_app}} - R^2_{\text{Nagelkerke}_{\text{adj}}} = \frac{R^2_{CS_app}}{\max(R^2_{CS})} - \frac{R^2_{CS_adj}}{\max(R^2_{CS})} \leq \delta \tag{6}$$

holds if the required level of shrinkage ($S_{VH}$) is such that:

$$S_{VH} \geq \frac{R^2_{CS_adj}}{R^2_{CS_adj} + \delta \max(R^2_{CS})}\,. \tag{7}$$

For binary logistic regression, the sample size for criterion (ii) is then calculated by substituting the minimum $S_{VH}$ that satisfies equation (7) into equation (4).

### 2.3. Overview of criterion (iii): ensure precise estimate of overall risk [12]

The third sample size criterion of Riley et al.[12] is to ensure a precise estimate of overall risk. For binary logistic regression, an approximate 95% confidence interval for the estimate of the overall outcome proportion ($\hat{\theta}$) can be expressed as:

$$\hat{\theta} \pm 1.96\sqrt{\frac{\hat{\theta}(1-\hat{\theta})}{n}}. \qquad (8)$$

Therefore, to target a pre-specified absolute margin of error of $\delta$, the following sample size is required:

$$n = \left(\frac{1.96}{\delta}\right)^2 \hat{\theta}\left(1-\hat{\theta}\right). \qquad (9)$$

The final (minimum) sample size is then taken to be the maximum sample size across criterion (i), (ii) and (iii).

# 3. Extending the sample size formula to multinomial logistic regression

In this section we extend each of the criteria from section 2 to the situation where the outcome has multiple categories, and we wish to develop a CPM using multinomial logistic regression. As such, hereto we consider an outcome, $Y_i$ $(i = 1, \ldots, N)$, which has $K$ nominal categories, where $Y_i = k$ for $k \in [1, K]$ if individual $i$ has the $k$-th outcome category. All other notation introduced in section 2 remains the same.

## 3.1. Introducing the multinomial logistic regression model and its calibration framework

A multinomial logistic regression model[24] predicting outcome, $Y_i$, with $K$ nominal categories (taking the first category as the reference, without loss of generality), and $Q$ the number of predictor parameters in each sub-model $k$, is expressed by the following set of equations (dropping the subscript $i$ for brevity):

$$P(Y=1)=\frac{1}{1+\exp\left(\beta_{0,2}+\sum_{q=1}^{Q}\beta_{q,2}X_q\right)+\cdots+\exp\left(\beta_{0,K}+\sum_{q=1}^{Q}\beta_{q,K}X_q\right)}$$
$$P(Y=2)=\frac{\exp\left(\beta_{0,2}+\sum_{q=1}^{Q}\beta_{q,2}X_q\right)}{1+\exp\left(\beta_{0,2}+\sum_{q=1}^{Q}\beta_{q,2}X_q\right)+\cdots+\exp\left(\beta_{0,K}+\sum_{q=1}^{Q}\beta_{q,K}X_q\right)} \quad (10)$$
$$\vdots$$
$$\vdots$$
$$P(Y=K)=\frac{\exp\left(\beta_{0,K}+\sum_{q=1}^{Q}\beta_{q,K}X_q\right)}{1+\exp\left(\beta_{0,2}+\sum_{q=1}^{Q}\beta_{q,2}X_q\right)+\cdots+\exp\left(\beta_{0,K}+\sum_{q=1}^{Q}\beta_{q,K}X_q\right)}$$

which equates to the following $K-1$ submodels:

$$\ln\left[\frac{P(Y=k)}{P(Y=1)}\right]=\beta_{0,k}+\sum_{q=1}^{Q}\beta_{q,k}X_q \quad (11)$$

for $k=2,\ldots,K$, alongside the constraint $\sum_{k=1}^{K}P(Y=k)=1$.

Sub-model specific shrinkage factors can be defined for a multinomial logistic regression based on the recalibration framework outlined by Van Hoorde et al.[25] Specifically, after fitting a multinomial logistic regression using maximum likelihood (equation 11), a separate shrinkage factor $S_{MN_k}$ is applied to all the $\beta$'s for each sub-model $k$, and each intercept updated (to ensure calibration-in-the-large), as follows:

$$\ln\left[\frac{P(Y=k)}{P(Y=1)}\right]=\alpha_{0,k}^{*}+S_{MN,k}\left(\sum_{q=1}^{Q}\hat{\beta}_{q,k}X_q\right) \quad (12)$$

for $k = 2, \dots, K$, where $\hat{\beta}_{m,k}$ are the maximum likelihood estimates from the multinomial logistic regression (equation 11), and $\alpha^*_{0,k}$ are the re-estimated intercepts.

## 3.2. Extending criterion (i) to multinomial logistic regression

### 3.2.1. Direct application of criterion (i) to multinomial logistic regression models

A natural starting point for criterion (i) from binary logistic regression to multinomial logistic regression, would be to again base the required sample on targeting a single heuristic shrinkage factor of van Houwelingen and Le Cessie[20] to be at or above the chosen threshold. If we let

$$S_{VH_MN} = 1 - \frac{(K-1) * Q}{LR_{MN}}\,, \qquad (13)$$

be the heuristic shrinkage factor of the multinomial model, where $LR_{MN} = -2(lnL_{null} - lnL_{model})$ is like likelihood ratio test statistic for the multinomial model, $(K-1) * Q$ is the total number of predictor parameters across all sub-models, and $R^2_{CS_adj} = 1 - \exp\left(\frac{-LR_{MN}}{n}\right)$ is the apparent estimate of the Cox-Snell generalised definition of $R^2$ for the multinomial model, then equation (4) could again be used to define a minimum required sample size to target $S_{VH_MN}$ to be at a pre-specified threshold:

$$n = \frac{(K-1)*Q}{\left(S_{VH_MN}-1\right)\ln\left(1-\frac{R_{CS_adj}^2}{S_{VH_MN}}\right)}. \quad (14)$$

However, this approach has issues. While the van Houwelingen and Le Cessie[20] heuristic shrinkage factor is an estimator of the (true) $S$ for binary logistic regression, there is no clear relationship between $S_{VH_MN}$ and the $K-1$ multinomial sub-model specific shrinkage factors ($S_{MN,k}$: $k = 2, \dots, K$) in equation (12). Therefore, simply ensuring that $S_{VH_MN}$ surpasses a pre-specified threshold (using equation 14) would not necessarily result in the required level of shrinkage in each sub-model (i.e., $S_{MN,k}$: $k = 2, \dots, K$ in equation (12)). This is undesirable because it would mean that some sub-models of the multinomial model could be overfit. Therefore, we propose another approach to extend criterion (i), which targets all sub-model shrinkage factors ($S_{MN,k}$: $k = 2, \dots, K$) to be at or above the desired threshold.

### 3.2.2. Alternative suggestion for criterion (i), utilising distinct logistic regression models

Equation (11) can be expressed as a set of $K-1$ distinct logistic regression models fitted separately, in the subset of the cohort which has either outcome $k$ or outcome 1 (the reference). That is, the following binary logistic regression model can be fitted,

$$\ln\left[\frac{P(Y=k)}{P(Y=1)}\right] = \gamma_{0,k} + \sum_{q=1}^{Q} \gamma_{q,k} X_q, \tag{15}$$

on the subset of individuals where $Y \in \{1, k\}$, separately for $k = 2, 3, \dots, K$. These are also referred to as "one vs one" models.[6]

Crucially, a separate shrinkage factor for each distinct logistic regression model can then be calculated,[26–28] such that

$$\ln\left[\frac{P(Y=k)}{P(Y=1)}\right] = \gamma_{0,k}^{*} + S_{DL,k}\left(\sum_{q=1}^{Q} \hat{\gamma}_{q,k} X_q\right), \tag{16}$$

where $\hat{\gamma}_{m,k}$ are the coefficients estimated from equation (15), $\gamma_{0,k}^{*}$ are the re-estimated intercepts, and $S_{DL,k}$ is the shrinkage factor for distinct logistic regression model $k$, defined in the same way as $S$ from section 2.1 (referred to as 'distinct logistic shrinkage factors'). This means that if an estimate of $R^2_{CS_adj}$ is available for each distinct logistic regression model $k$, then using the process summarised in section 2.1, equation (4) can be used to derive a sample size to target a particular distinct logistic shrinkage factor for each model. Importantly, it has been shown that the sub-models of the multinomial logistic regression, and the distinct logistic regression models, are parametrically equivalent ($\gamma_{m,k} = \beta_{m,k}$).[29] Given the asymptotically unbiased property of the maximum likelihood estimators, it follows that $\hat{\gamma}_{m,k} = \hat{\beta}_{m,k}$ as $N \to \infty$, and hence $S_{DL,k} \to S_{MN,k}$ as $N \to \infty$. Therefore, deriving a sample size to target the shrinkage factor of each distinct logistic regression ($S_{DL,k}$: $k = 2, \ldots, K$) is above the desired value using equation (4),[12] will also target the multinomial sub-model specific shrinkage factors ($S_{MN,k}$: $k = 2, \ldots, K$) to be above the desired threshold. A separate sample size calculation must therefore be done for each pair of outcomes, taking the maximum to ensure criterion (i) is satisfied for each sub-model.

### 3.2.3. Consideration of the choice of reference category

So far, the first outcome category has been taken as the reference. During model development, changing the reference category will not have an impact on the risk scores generated from the model. However, upon validating a multinomial CPM, the choice of reference category will change which of the multinomial sub-model specific shrinkage factors are calculated (i.e., what is estimated from equation (12)). Therefore, for the purposes of the criterion (i), one must ensure that the shrinkage of every possible sub-model across all reference categories at model validation is above a certain level. Not doing so (i.e., only focusing on one outcome category), would potentially create over-confidence in the model's ability to distinguish between some of the outcome categories. In other words, one must aim to minimise optimism in all pairwise performance metrics. While calculating criterion (i) with taking each outcome category as reference in turn may lead to high required sample sizes (e.g., see section 5), this would be reflective of one trying to develop a CPM to predict multinomial outcomes that require a lot of statistical power; this should be viewed as valuable information rather than a hindrance (as with any sample size calculation). We therefore outline our final approach in the following section, to ensure overfitting is minimised across all pairs of outcomes.

### 3.2.4. Final proposal for criterion (i)

Let $S_{MN,k,r}$ be the sub-model specific shrinkage factor from the van Hoorde et al.[25] framework, for sub-model $k$ with the reference category $r$. The following approach will target every $S_{MN,k,r}$ to be above the pre-specified threshold. The proposal is to follow the approach outlined in section 3.2.2 for every possible reference category, and take the maximum sample size across all reference categories. That is for each distinct logistic regression model $\{k, i\}$ where $k \neq r$,

$$\ln\left[\frac{P(Y=k)}{P(Y=r)}\right] = \gamma_{0,k,r} + \sum_{q=1}^{Q} \gamma_{q,k,r} X_q, \qquad (17)$$

we can obtain a corresponding shrinkage factor, $S_{DL,k,r}$, each defined in the same way as $S_{DL,k}$ from equation (16); that is,

$$\ln\left[\frac{P(Y=k)}{P(Y=r)}\right] = \gamma^*_{0,j,r} + S_{DL,k,r}\left(\sum_{q=1}^{Q} \hat{\gamma}_{q,k,r} X_q\right). \qquad (18)$$

Define $m_{k,r}$ as the number of individuals with outcome category $k$ or $r$ that is required to target the shrinkage factor $S_{DL,k,r}$ to be above some pre-defined threshold (e.g., 0.9). Here, $m_{k,r}$ can be calculated using the existing formula for binary logistic regression[12]; specifically, using equation (4). To do so, appropriate estimates of $R^2_{CS,k,r}$ (Cox-Snell $R^2$ for distinct logistic regression model $\{k, r\}$), $Q$ (the number of predictor parameters considered for inclusion in each sub-model), and $p_k$ (the proportion of individuals from the cohort expected to have outcome $k$) must each be pre-specified. Suggestions of how to pre-specify $R^2_{CS,k,r}$ are given in section 4. Note that because the logistic regression models for $\ln\left[\frac{P(Y=k)}{P(Y=r)}\right]$ and $\ln\left[\frac{P(Y=r)}{P(Y=k)}\right]$ are equivalent, we can reduce this to only consider the combinations where $k > r$. From now on when we refer to sub-models, we are referring to all sub-models $\{k, r\}$, where $k > r$ (there are $0.5 * K * (K-1)$ of these). An exception to this is when discussing model development, for example the number of predictor parameters $Q$ considered for inclusion in each sub-model. We assume that the same set of variables are considered for inclusion in each sub-model and the same final set of variables will be included in each sub-model, which is common in practice. In this situation, changing the reference category will not impact the risk scores, and therefore we are effectively referring to any $K-1$ sub-models sharing a reference category.

The total number of individuals required in the whole cohort ($n_{k,r}$) to ensure there are $m_{k,r}$ individuals with outcome categories $\{k, r\}$, can then be calculated as $n_{k,r} = m_{k,r}/p_{k,r}$ ,where $p_{k,r}$ is the proportion of individuals from the whole cohort expected to have outcome categories $\{k, r\}$. Finally, the required sample size $n$ to satisfy our criterion (i) is taken to be $n = max(n_{k,r}: k > r)$.

The proposed approach to implementing criterion (i) are evaluated in a simulation study with full details provided in Appendix S1, (section 1).

## 3.3. Extending criterion (ii) to multinomial logistic regression

As noted in section 2.2, the second criterion outlined by Riley et al.,[12] is defined to ensure a small difference $(\delta)$ between the observed and expected proportion of variance explained ($R^2_{\text{Nagelkerke}}$) for the overall model.

As outlined in de Jong et al.,[16] the apparent $R^2_{\text{Nagelkerke}}$ for a multinomial logistic regression model is defined in the same as for a binary logistic model:

$$R^2_{\text{Nagelkerke}} = \frac{1 - \exp\left(\frac{-LR_{MN}}{n}\right)}{1 - \exp\left(\frac{2lnL_{null}}{n}\right)} = \frac{R^2_{CS}}{\max\left(R^2_{CS_app}\right)} \ , \qquad (19)$$

where $LR_{MN}$ is as defined previously, and $lnL_{null}$ the log-likelihood of an intercept only multinomial model. Therefore, given the definition of $S_{VH_MN}$ in equation (13), to ensure a difference of less than $\delta$ between the apparent and adjusted $R^2_{\mathrm{Nagelkerke}}$ the following equation must hold:

$$S_{VH_MN} \geq \frac{R^2_{CS_adj}}{R^2_{CS_adj} + \delta \max\left(R^2_{CS_app}\right)}\,, \qquad (20)$$

Plugging this into equation (14), and noting that $n$ is a monotonically increasing function of $S_{VH_MN}$, we get the following requirement for criterion (ii) for multinomial logistic regression prediction models:

$$n \geq \frac{(K-1) * Q}{\left(\frac{R^2_{CS_adj}}{R^2_{CS_adj} + \delta \max\left(R^2_{CS_app}\right)} - 1\right) \ln\left(1 - R^2_{CS_adj} - \delta \max\left(R^2_{CS_app}\right)\right)}\,, \qquad (21)$$

Given $R^2_{\mathrm{Nagelkerke}}$is similarly defined for binary logistic and multinomial logistic regression models, this criterion is directly transferable from binary logistic regression to multinomial models. In line with the criterion (section 2.2) for binary logistic regression, we recommend a difference of $\delta$ = 0.05.[11,12]

We note that for criterion (ii), we focus on the fit of the overall multinomial logistic regression model, in contrast to criterion (i) where we focused on each sub-model. The reason for this is that $R^2_{CS}$ (and hence $R^2_{\text{Nagelkerke}}$) is not typically expressed for the sub-models of a multinomial logistic regression. While we could ensure that criterion (ii) holds for each distinct logistic regression model, it is not clear what this would achieve with respect to the sub-models of the multinomial logistic regression model.

## 3.4. Extending criterion (iii) to multinomial logistic regression

As outlined in section 2.3, the third criterion of Riley et al.,[12] is to ensure a precise estimate of overall risk (i.e., model intercept). To mimic the approach for binary logistic regression, for a multinomial model, this can be done by calculating the margin of error in the outcome proportion estimates.

Let $p_k = E_k/n$ be the proportion of individuals from the entire cohort with outcome category $\{k\}$, with $E_k$ the number of events in outcome category $k$. If $\pi_k$ is the underlying multinomial probability of outcome category $k$, then it can be shown through the work of Quesenberry and Hurst[30], and Goodman[31], that the simultaneous $\alpha \times 100\%$ confidence interval limits for $\pi_1, \pi_2, \dots \pi_K$:

$$\pi_k^- \leq \pi_k \leq \pi_k^+; k = 1, \dots, K$$

can be estimated by

$$p_k \pm \left[\chi^2_{\frac{\alpha}{K},1} \times \frac{p_k(1-p_k)}{n}\right]^{\frac{1}{2}} \quad (22)$$

Where , 1 denotes the Chi-squared distribution with 1 degree of freedom. Therefore, the sample size to ensure an absolute margin of error $\delta$ (say 0.05) at a $(1-\alpha) \times 100\%$ confidence level is

$$n = \max_{k=1,\dots,K}\left(\frac{\chi^2_{\frac{\alpha}{K},1} \times p_k(1-p_k)}{\delta^2}\right). \quad (23)$$

We choose to target simultaneous confidence intervals[30,31] rather than pointwise confidence intervals, so that every estimate of overall risk will simultaneously be within the pre-defined margin of error. This will require a larger sample size than considering pointwise confidence intervals, and is therefore conservative.

A summary of our proposed sample size criteria for a multinomial logistic regression model is given in box 1.

**STEP 1: Choose number of predictor parameters $Q$ considered for inclusion in each sub-model at model development**

Recognise that one predictor may require > 1 predictor parameter; for example, categorical predictor with > 2 categories, a continuous predictor with nonlinear terms, and interaction terms.

**STEP 2: Choose sensible values for $p_k$, the proportion of individuals in the cohort with outcomes in category $k$, $\max(R^2_{CS_app})$ and $R^2_{CS_adj}$ for the multinomial model, and $R^2_{CS_adj,k,r}$ of each distinct logistic regression models**

Ideally this will be based on previously published models in the same setting with similar outcome definition, a variety of ways to estimate these from various reported statistics are given in section 4. If no previous information is available to estimate $R^2_{CS_adj,k,r}$, use values which correspond to an $R^2_{\text{Nagelkerke}} = 0.15$ in each sub-model.

**STEP 3: Criterion (i)**

1. Calculate the minimum sample size ($m_{k,r}$) for each distinct logistic regression model $\{k, r\}$, where $k > r$, using equation (4) based on a pre-specified level of shrinkage (for example, targeting shrinkage factors of 0.9) and an estimate of $R^2_{CS_adj,k,r}$.

2. Calculate the total number of individuals needed to achieve the required number in each distinct logistic regression model $\{k, r\}$, by dividing by $p_{k,r}$, $n_{k,r} = m_{k,r}/p_{k,r}$

3. Take the minimum sample size for criterion (i) to be $n = max(n_{k,r}: k > r)$, which will target all the multinomial sub-model specific shrinkage factors to be greater-than-or-equal to the pre-specified threshold.

**STEP 4: Criterion (ii)**

Use equation (21) to calculate a sample size to target the difference between the apparent and optimism adjusted $R^2_{\text{Nagelkerke}}$ to be $\delta$, using estimates of $\max(R^2_{CS_app})$ and $R^2_{CS_adj}$. Previously $\delta = 0.05$ has been recommended.[14]

**STEP 5: Criterion (iii)**

Use equation (23) to calculate a sample size to target the simultaneous 95% confidence intervals of the estimates of overall risk for each category to be $\leq \delta$, using estimates of $p_k$. We recommend $\delta = 0.05$.

**STEP 6: FINAL SAMPLE SIZE**

The required minimum sample size is the maximum value from steps 3 to 5, to ensure criteria (i), (ii) and (iii) are met.

*Box 1: Summary of the proposed minimum sample size criteria for*

*multinomial logistic regression CPMs*

# 4. Practical recommendations for implementing criteria in practice (estimating $\boldsymbol{R^2_{CS_adj}}$ and dealing with large required sample sizes)

To perform our proposed sample size calculations, an estimate of $R^2_{CS_adj}$ needs to be pre-specified. As with earlier work[11,12,14] we recommend that this is based on similar, previously developed or validated predictions models. When calculating criterion (i), estimates of $R^2_{CS_adj,k,r}$ are required for each distinct logistic regression model $\{k, r\}$, corresponding to the sub-models of the multinomial logistic model. When calculating criterion (ii) an estimate of $R^2_{CS_adj}$ is required for the multinomial logistic regression model. We discuss how to estimate these using published data below. We also urge researchers to report the metrics discussed below when publishing future CPM development papers based on multinomial logistic regressions, to aid the sample size calculations of others. Finally we make recommendations on what to do if the calculated required sample size is unfeasibly high.

## 4.1. Recommendations for deriving $R^2_{CS_{adj},k,i}$ of distinct logistic regression models

To calculate criterion (i) estimates of $R^2_{CS_adj,k,r}$ are required. If the appropriate “one-vs-one”[6] distinct logistic regression models have been fitted in a published study and estimates of $R^2_{CS_adj,k,r}$ have been reported, these can be used directly. If other pseudo-$R^2$ statistics have been reported (for each distinct logistic), there are a variety of ways to derive $R^2_{CS_adj,k,r}$ from these; see Riley et al.[12] Alternatively, if the C-statistics of each distinct logistic regression model are available, then $R^2_{CS_adj,k,r}$ can be estimated using a simulation approach.[15]

However, it is highly likely that each distinct logistic regressions will not have been fitted alongside any previously developed multinomial logistic regression model. In this case, the pairwise C-statistics[32] (using the conditional risk method) of the multinomial logistic regression might have been reported. Here, since these pairwise C-statistics provide an estimate of the C-statistic for each distinct logistic regression model, they can be used to estimate $R^2_{CS_adj,k,r}$ using the simulation approach of Riley et al.[15] We illustrate this approach in our worked example in section 5.

If neither pseudo-$R^2$ or (pairwise) C-statistics are available a priori, we suggest calculating the minimum sample size following the approach suggested by Riley et al.,[14] for when information on $R^2_{CS_adj}$ is not available. Specifically, under a conservative assumption of optimism adjusted $R^2_{\text{Nagelkerke}}$ of 0.15 (15%), equation (5) can be modified to give $R^2_{CS_app,k,r} = 0.15 * \max{(R^2_{CS_app,k,r})}$ for each distinct logistic regression model. Here, $\max{(R^2_{CS_app,k,r})}$ can be estimated for each model using equation (6):

$$\max\left(R^2_{CS_app,k,r}\right) = 1 - \exp\left(\frac{2 * lnL_{null,k,r}}{n}\right), \qquad (24)$$

where $lnL_{null,k,r}$ can be calculated for each distinct logistic regression models using:

$$lnL_{null,k,r} = E_k \ln\left(\frac{E_k}{E_k + E_r}\right) + E_r \ln\left(\frac{E_r}{E_k + E_r}\right), \qquad (25)$$

where $E_k$ and $E_r$ are the number of outcome events in category $k$ and $r$ respectively. Alternatively (and equally), for each distinct logistic regression model $\max{(R^2_{CS_app,k,r})}$ can be calculated as:

$$\max\left(R^2_{CS_app,k,r}\right) = 1 - \left({\varphi_{k,r}}^{\varphi_{k,r}}\left(1 - \varphi_{k,r}\right)^{1-\varphi_{k,r}}\right)^2, \qquad (26)$$

where $\varphi_{k,r} = E_k/E_k + E_r$, is the outcome proportion in category $k$ relative to the reference category $r$. If a multinomial model had been published, then this information would be available for each distinct logistic regression model assuming the number of events in each category had been reported.

## 4.2. Recommendations for deriving $R^2_{CS_adj}$ of multinomial logistic regression models

To calculate criterion (ii) a pre-specified estimate of the overall $R^2_{CS_adj}$ is required. As previous, this would ideally be based on information from a previous multinomial logistic regression model. Similarly to binary logistic regression, if other pseudo-$R^2$ statistics have been reported, there are a variety of ways to derive $R^2_{CS_adj}$ from these, as outlined in Riley et al.[12]

Alternatively, one could again take a conservative approach of setting $R^2_{CS_adj} = 0.15 * \max\,(R^2_{CS_app})$ (corresponding to an $R^2_{\text{Nagelkerke}}$ of 0.15). There are two ways to calculate $\max\left(R^2_{CS_app}\right)$ for multinomial logistic regression. The first is to use equation (6), where $lnL_{null}$ can be calculated for a multinomial logistic regression as:

$$lnL_{null} = \sum_{k=1}^{K} E_k \ln\left(\frac{E_k}{n}\right), \qquad (27)$$

with $E_k$ denoting the number of events in outcome category $k$.

Alternatively, $\max(R^2_{CS_app})$ can be expressed as:

$$\max(R^2_{CS_app}) = 1 - \left(\prod_{k=1}^{K} (p_k)^{p_k}\right)^2, \qquad (28)$$

where $p_k = E_k/n$ is the is the observed frequency of category $k$, as defined in section 3.4. This expression follows naturally from equations (6) and (27), and details of its derivation are given in Appendix S1 (section 3). Some implications of basing the estimate of $R^2_{CS_app}$ on the assumption $R^2_{\text{Nagelkerke}} = 0.15$ are given in Appendix S1 (section 4).

## 4.3. Recommendations if required sample size is too high

We propose three strategies if the required sample size is completely unfeasible to recruit. It is worth reiterating, that the estimated sample size is required to build the proposed model with the specified levels of overfitting, optimism and precision. In order to reduce the sample size, the model must either be simplified, or you must be willing to accept overfitting, optimism and precision below the desired level.

1) Merge outcome categories. We believe the first consideration could be to merge some of the outcome categories that are driving the high sample size; looking at each pairwise criterion (i) will indicate which categories are driving the sample size. This decision would have to be made alongside clinical considerations, such as which outcome categories could be merged whilst retaining clinical interpretation.

2) Reduce the number of predictor parameters. A second suggestion is to reduce the number of predictor parameters, which is inline with previous suggestions.[11,12,14] However, we have only looked at scenarios where there are a fixed number of predictor parameters considered for each sub-model. This means when reducing the number of predictor parameters, one would be doing so across all sub-models. An alternative to this is to only reduce the number of predictor parameters considered for inclusion in the sub-model(s) with the highest level of overfitting. This is an enticing approach, as one does not want to reduce the number of predictor parameters in sub-models which are not suffering from overfitting. However, the implications of such an approach are not yet clear and would require further research before this could be recommended.

3) Reduce the acceptable level of overfitting between specific pairs of outcomes. Rather than having the acceptable level of shrinkage at 0.9, it could be reduced (e.g., to 0.8), specifically for the pair of outcomes which are driving the high sample size. This is somewhat undesirable as criterion (i) is in place to minimise overfitting. However, at least the targeted level of overfitting would be explicitly stated and the limitations of the model would therefore be well quantified.

# 5. A worked example of calculating sample size criteria for a multinomial logistic regression model

## 5.1. Hypothetical scenario and information available in literature

In this section, we present a worked example to illustrate how our proposed sample size criteria could be implemented in practice. The code that was used to do this is available on GitHub.[33] Our example aims to calculate the minimum sample size required to develop a multinomial logistic regression prediction model to predict the tumour type (benign, borderline, stage I invasive, stage II-IV invasive or metastatic) when presented with an ovarian mass. This is an important preoperative diagnosis, as dependent on the type of tumour, different clinical action may be taken.

Van Calster et al.[8] considered the development of such a model using the International Ovarian Tumor Analysis Group[34] dataset. The following information is available from that work. The model was developed on a dataset of 3506 tumours, of which 2557 were benign, 186 were borderline, 176 were stage I invasive, 467 were stage II-IV invasive, and 120 were metastatic. The following pairwise C-statistics[32] were reported for every combination of outcome comparisons: 0.85 (benign vs borderline), 0.92 (benign vs stage I invasive), 0.99 (benign vs stage II-IV invasive), and 0.95 (benign vs metastatic), 0.75 (borderline vs stage I invasive), 0.95 (borderline vs stage II – IV invasive), 0.87 (borderline vs metastatic), 0.87 (stage I invasive vs stage II – IV invasive), 0.71 (stage I invasive vs metastatic) and 0.82 (stage II – IV invasive vs metastatic). These pairwise C-statistics are reported from a temporal validation and are free from in-sample optimism concerns, therefore we can use these to estimate $R^2_{CS_adj,k,i}$ directly with no adjustment for optimism required. There were 17 candidate predictor parameters considered for inclusion in the model, including all the fractional polynomials of continuous variables. We will assume we will consider the same set of variable for inclusion in this worked example before applying variable selection techniques. We now illustrate the use of the aforementioned information to perform our sample size calculation.

## 5.2. Steps 1 and 2: Identifying values for , $p_k$, $p_{k,r}$, $\max\left(R^2_{CS_app}\right)$, $R^2_{CS_adj}$ and $R^2_{CS_adj,k,r}$.

$k$ and $r$ take the values 1 (benign), 2 (borderline), 3 (stage I invasive), 4 (stage II-IV invasive) and 5 (metastatic).

### 5.2.1. Calculating $Q$

Assuming we consider the same set of variables for variable selection that were used in the work by Van Calster et al.,[8] this would mean $Q = 17$.

### 5.2.2. Calculating $p_k$ and $p_{k,r}$

$p_k$ is the proportion of individuals that have outcome category $\in \{k\}$, $p_{k,r}$ is the proportion of individuals that have outcome category $\in \{k, r\}$ where $k > r$. To estimate these values, we use the prevalence of each outcome category as reported in section 5.1: $p_1 = 0.729$, $p_2 = 0.053$, $p_3 = 0.050$, $p_4 = 0.133$, $p_5 = 0.034$, $p_{2,1} = 0.782$, , $p_{3,1} = 0.780$, $p_{4,1} = 0.863$, $p_{5,1} = 0.764$, $p_{3,2} = 0.103$, $p_{4,2} = 0.186$, $p_{5,2} = 0.087$, $p_{4,3} = 0.183$, $p_{5,3} = 0.084$, $p_{5,4} = 0.167$.

### 5.2.3. Calculating $\max(R^2_{CS_app})$

We calculated $\max(R^2_{CS_app})$ using equation (28), and the prevalence of each outcome category $p_k$:

$$\max\left(R^2_{CS_{app}}\right) = 1 - (0.729^{0.729} * 0.053^{0.053} * 0.050^{0.050} * 0.133^{0.133} * 0.034^{0.034})^2 = 0.841.$$

### 5.2.4. Calculating $R^2_{CS_adj}$

Given the $R^2_{CS_adj}$ of the overall multinomial model had not been reported, we based our estimate of $R^2_{CS_adj}$ on assuming $R^2_{\text{Nagelkerke}} = 0.15$. Using the estimate of $\max(R^2_{CS_app})$ in equation (5) gave an estimate of:

$$R^2_{CS_adj} = 0.15 * \max\left(R^2_{CS_{app}}\right) = 0.15 * 0.841 = 0.126\,.$$

### 5.2.5. Calculating $R^2_{CS_adj,k,r}$

No data was available on the $R^2_{CS_adj,k}$ in the work of Van Calster et al.,[8] therefore we had to estimate them indirectly using the simulation approach of Riley et al.[15] This method utilises the pairwise C-statistics, on which data was available.[8] Estimates of the pairwise outcome proportions $\varphi_{k,r}$ of category $k$ relative to the reference category $r$ are also required to implement the simulation approach. These were estimated using the number of tumours in each category: $\varphi_{2,1} = 0.068$, $\varphi_{3,1} = 0.064$, $\varphi_{4,1} = 0.154$, $\varphi_{5,1} = 0.045$, $\varphi_{3,2} = 0.486$, $\varphi_{4,2} = 0.715$, $\varphi_{5,2} = 0.392$, $\varphi_{4,3} = 0.726$, $\varphi_{5,3} = 0.405$, $\varphi_{5,4} = 0.204$. The simulation approach was followed for each sub-model to give estimates of: $R^2_{CS_adj,2,1} = 0.116$, $R^2_{CS_adj,3,1} = 0.179$, $R^2_{CS_adj,4,1} = 0.497$, $R^2_{CS_adj,5,1} = 0.170$, $R^2_{CS_adj,3,2} = 0.185$, $R^2_{CS_adj,4,2} = 0.499$, $R^2_{CS_adj,5,2} = 0.374$, $R^2_{CS_adj,4,3} = 0.328$, $R^2_{CS_adj,5,3} = 0.129$, $R^2_{CS_adj,5,4} = 0.210$. Code for this simulation approach is provided at the GitHub repsitory,[33] as well as more general code on how to implement this simulation approach in the original work.[15]

## 5.3. Step 3: Criterion (i)

Following the process in section 3.2.4, first each $m_{k,r}$ was calculated using equation (4) and the estimates of $R^2_{CS_adj,k,r}$ from section 5.2.5. Then the total number of individuals required to target a multinomial sub-model specific shrinkage factor of 0.9 for sub-model $\{k, r\}$, $n_{k,r}$, was calculated by dividing $m_{k,r}$ by $p_{k,r}$:

$$n_{2,1} = m_{2,1}/p_{2,1} = \frac{17}{(0.9-1)\ln\left(1-\frac{0.116}{0.9}\right)} * \frac{3506}{2557+186} = 1574$$

$$n_{3,1} = m_{3,1}/p_{3,1} = \frac{17}{(0.9-1)\ln\left(1-\frac{0.179}{0.9}\right)} * \frac{3506}{2557+176} = 982$$

$$n_{4,1} = m_{4,1}/p_{4,1} = \frac{17}{(0.9-1)\ln\left(1-\frac{0.497}{0.9}\right)} * \frac{3506}{2557+467} = 246$$

$$n_{5,1} = m_{5,1}/p_{5,1} = \frac{17}{(0.9-1)\ln\left(1-\frac{0.170}{0.9}\right)} * \frac{3506}{2557+120} = 1067$$

$$n_{3,2} = m_{3,2}/p_{3,2} = \frac{17}{(0.9-1)\ln\left(1-\frac{0.185}{0.9}\right)} * \frac{3506}{186+176} = 7147$$

$$n_{4,2} = m_{4,2}/p_{4,2} = \frac{17}{(0.9-1)\ln\left(1-\frac{0.499}{0.9}\right)} * \frac{3506}{186+467} = 1128$$

$$n_{5,2} = m_{5,2}/p_{5,2} = \frac{17}{(0.9-1)\ln\left(1-\frac{0.374}{0.9}\right)} * \frac{3506}{186+120} = 3629$$

$$n_{4,3} = m_{4,3}/p_{4,3} = \frac{17}{(0.9-1)\ln\left(1-\frac{0.328}{0.9}\right)} * \frac{3506}{176+467} = 2045$$

$$n_{5,3} = m_{5,3}/p_{5,3} = \frac{17}{(0.9-1)\ln\left(1-\frac{0.129}{0.9}\right)} * \frac{3506}{176+120} = 13063$$

$$n_{5,4} = m_{5,4}/p_{5,4} = \frac{17}{(0.9-1)\ln\left(1-\frac{0.210}{0.9}\right)} * \frac{3506}{467+120} = 3813$$

The minimum required sample size was taken as the maximum of these, and therefore $N = 13063$, approximately 9527 benign tumours, 693

borderline, 656 stage I invasive, 1740 stage II-IV invasive and 447 metastatic (assuming same outcome proportions as in Van Calster et al.[8]).

## 5.4. Step 4: Criterion (ii)

Criterion (ii) aims to calculate a sample size required to ensure a difference of 0.05 between the apparent and adjusted $R^2_{\text{Nagelkerke}}$, which holds if equation (21) is satisfied (section 3.3). Plugging in the estimates of $\max\left(R^2_{CS_app}\right)$ and $R^2_{CS_adj}$ into equation (21) gives:

$$n \geq \frac{4 * 17}{\left(\frac{0.126}{0.126 \ + 0.05 * 0.841} - 1\right) \ln(1 - 0.126 - 0.05 * 0.841)}$$

$$n \geq 1477 \ .$$

So 1477 individuals are required to meet criterion (ii), approximately 1077 benign tumours, 78 borderline, 74 stage I invasive, 197 stage II-IV invasive, and 51 metastatic.

## 5.5. Step 5: Criterion (iii)

Criterion (iii) is to ensure a precise estimate of risk in the overall population. Following the steps outlined in section 3.4, for a 95% confidence interval ($\alpha = 0.05$), using the estimated values for $p_k$, with $K = 5$ and an absolute margin of error of $\delta = 0.05$, then the required sample size for each outcome is (equation 21):

$$n_1 = \frac{6.635 \times 0.729(1 - 0.729)}{0.05^2} = 524$$
$$n_2 = \frac{6.635 \times 0.053(1 - 0.053)}{0.05^2} = 134$$
$$n_3 = \frac{6.635 \times 0.50(1 - 0.150)}{0.05^2} = 127$$
$$n_4 = \frac{6.635 \times 0.133(1 - 0.133)}{0.05^2} = 307$$
$$n_5 = \frac{6.635 \times 0.034(1 - 0.034)}{0.05^2} = 88$$

Leaving the sample size for criterion (iii) to be $\max(n_1, n_2, n_3, n_4, n_5) = 524$, approximately 382 benign tumours, 28 borderline, 26 stage I invasive, 70 stage II-IV invasive, and 18 metastatic.

## 5.6. Step 6: Final sample size

The final sample size is taken as the maximum of the sample sizes required to satisfy each criterion (i)-(iii), which was 13063 (i), 1476 (ii) and 524 (iii) respectively. Hence, the minimum required sample size would be $n = 13063$, approximately 9527 benign tumours, 693 borderline, 656 stage I invasive, 1740 stage II-IV invasive and 447 metastatic (assuming same outcome proportions as in Van Calster et al.[8]). For contrast, using the definition of events per variable ($EPV_m$) from De Jong et al.,[16] an $EPV_m$ of $10$ would result in a sample size of $4967$, and an $EPV_m$ of $20$ would result in a sample size of $9934$.

## 5.7. Suggestions for dealing with high sample size

The required sample size is high and is being driven by outcome categories 3 (stage I invasive) and 5 (metastatic). If the proposed model was developed with a sample size smaller than 13063, the level of overfitting between these two outcomes would not be targeted at the pre-specified level of $0.9$. Following the suggestions in section 4.3, the first solution would be to merge categories 3 and 4 (stage I invasive with stage II – IV invasive). With such a combination the model would retain clinical interpretation. If it was essential to keep these outcome categories separate, fewer predictor parameters could be considered instead. The value of $Q = 17$ incorporates fractional polynomial terms and interactions which could be removed, or one of the predictors could be removed altogether. A final possible option is to reduce the targeted level of overfitting for pair $\{3,5\}$. Plugging a value of $0.8$ into section 5.3 would give $n_{5,3} = 5746$, and the final sample size would be driven by $n_{3,2} = 7147$. While this is slightly undesirable, the targeted level of overfitting for all other outcome pairs would still be at $0.9$, and one could report that overfitting may be more likely for outcome pair $\{3,5\}$.

## 6. Discussion

We have presented sample size criteria for the development of prediction models for multiple-category outcomes using multinomial logistic regression. This builds upon recent developments in this space for continuous, binary and time-to-event outcomes.[11,12] Criterion (ii) and (iii) both had a natural extension into a multinomial framework. Criterion (i) did not and therefore we tested the properties of our proposed approach in a simulation (Appendix S1: section 1), finding that the sample size resulted in the desired level of overfitting. Our approach to criterion (i) may lead to high sample sizes if some of the outcome categories are rare, or have a low pairwise $R^2_{CS}$, however this is necessary if you want to ensure overfitting is minimised in prediction between all pairs of outcome categories. If the required sample size cannot be achieved, we have made some recommendations on how the model could be adjusted to lower the number required (section 4.3).

The biggest practical challenge with implementing these recommendations in practice is the availability of information on past $R^2_{CS}$, given multinomial logistic regression CPMs are not (yet) very common. The proposed criteria will be most effective in achieving their aim when an accurate estimate of the $R^2_{CS}$ is available for both the multinomial model ($R^2_{CS_adj}$) and each distinct logistic regression model $\{k, r\}$ ($R^2_{CS_adj,k,r}$). We have given advice on how to pre-specify these (sections 4.1 and 4.2), but also want to urge researchers to report the relevant information when developing a multinomial logistic regression to enable this process. Currently there is no way to estimate $R^2_{CS_adj}$ for the multinomial model from metrics which are not pseudo-$R^2$ (for example there is no way to estimate it from the PDI[35]), meaning reporting $R^2_{CS_adj}$ is very important. Estimates of $R^2_{CS_adj,k,r}$ can be obtained from a variety of metrics from previously published "one-to-one" distinct logistic regression models (section 4.1).[12,15] However, when fitting a multinomial logistic regression, it is important to (at least) report the pairwise C-statistics using the conditional risk method[32] when fitting a multinomial logistic regression model. This is an informative performance metric that should be reported anyway, and it will allow future researchers to estimate $R^2_{CS_adj,k,r}$ (as was done in our worked example). In theory, the conditional risk method could also be used to report $R^2_{CS_adj,k,r}$ directly, although we are not aware of any instances of people doing this is in the literature.

There are five important areas of future work. First, to establish a relationship between the PDI,[35] a commonly reported statistic for discrimination of a multinomial logistic regression model, and the distribution of the linear predictors of the sub-models. This relationship has been established for the C-statistic and logistic regression[36–39] allowing $R^2_{CS}$ to be estimated when only the C-statistic is available.[15] Secondly, the simulation (Appendix S1, section 1) found that there was poor agreement between the heuristic shrinkage factors and the sub-model specific shrinkage factors when covariate effects were large and sample sizes were small (Table S3). This resulted in not having the desired level of shrinkage in the developed models. This finding extends to binary logistic regression, but it is not clear whether similar results would be found for continuous or time-to-event outcomes. Given the proposed criterion (i) for every outcome type[11,12] targets the heuristic shrinkage factor to be at the chosen threshold, it's important to establish in which scenarios where this may be a poor predictor of the sub-model specific shrinkage factors. Third, to extend the criteria of van Smeden et al.,[13] to multinomial logistic regression. Their work acts as a fourth criterion,[14] to target the mean absolute prediction error (MAPE) of a binary logistic regression model to be below a pre-specified threshold. This helps ensure precise predictions across the spectrum of predicted values.  The formula is derived from a detailed simulation, in which a variety of binary logistic regression models are simulated and the MAPE assessed when the model is applied to new individuals from the target population. The first step to extending this criteria to multinomial logistic regression would be to define an extension of the MAPE for multinomial outcomes, which would then need to be followed by a similar simulation as the one used to derive the formula for binary logistic regression. The fourth is to develop sample size formula for the prediction of ordinal outcomes. The sample size formula proposed in this study are for a multinomial logistic regression, which can be fit to either nominal or ordinal outcomes. However if wanting to predict an ordinal outcome, an ordinal model could be fitted which would likely require a smaller sample size since they require less parameters to be estimated. The fifth is to explore the idea of reducing the number of predictor parameters in specific sub-models of the multinomial logistic regression as a way to reduce the required sample size, as discussed in section 4.3.

These sample size criteria will be embedded into existing software (pmsampsize in R[40] and Stata[41]) so they can be widely implemented in practice.

# 8. Supporting Statements

## 8.1. Acknowledgements

None

## 8.2. Funding

This work was supported by funding from the MRC-NIHR Methodology Research Programme [grant number: MR/T025085/1]. GSC was supported by the NIHR Biomedical Research Centre, Oxford, and Cancer Research UK (programme grant: C49297/A27294).

## 8.3. Availability of data and materials

Full reusable code for running the simulation and worked example are provided at the referenced GitHub repoistory.[33]

## 8.4. Author contributions

**AP, GM** and **RR** conceived and designed the study. **AP** conducted the analysis and interpreted the results in discussion with **GM** and **RR**. **AP** wrote the initial draft of the manuscript with support from **GM**, which was then critically reviewed for important intellectual content by **GM**, **RR**, **GSC**, **MVS**, **JE** and **BVC**. **JE** will embed the methods into R and Stata software. All authors have approved the final version of the paper.

## 8.5. Competing interests

None

## 8.6. Ethics approval

Ethical approval was not required for this study.

# Supplementary Material - Minimum Sample Size for Developing a Multivariable Prediction Model using Multinomial Logistic Regression

Alexander Pate, PhD, Richard D Riley, PhD, Gary S Collins, PhD, Maarten van Smeden, PhD, Ben Van Calster, PhD, Joie Ensor, PhD, Glen P Martin, PhD

## Contents

# 9. Full details of simulation study investigating the properties of the proposed alternative to criterion (i)

## 9.1. Methods

Methods reported using the ADEMP structure.[1] The simulation was carried out in R[2] version 3.4.1, using packages VGAM,[3] foreach[4] and doParallel.[5] Reproducible code is provided at the referenced GitHub repository.[6]

### 9.1.1. Simulation aims

A simulation study was carried out that evaluated the performance of the suggested ways to implement criterion (i) for multinomial logistic regression outlined in sections 3.2.1 and 3.2.2. Our primary aim was to investigate whether the sub-model specific shrinkage factors from the multinomial calibration framework were greater-than-or-equal to the pre-defined threshold used in the sample size calculations. Note that criterion (i) aims to keep the sub-model specific shrinkage factors above the pre-defined threshold *on average*, therefore we are interested in whether the mean and median of the sub-model specific shrinkage factors across all the simulations are above the threshold, rather than in every simulated dataset. Secondly, the simulation investigated whether the sub-model specific shrinkage factors in the multinomial logistic regression matched the distinct logistic shrinkage factors, to test the theory behind our proposed solution outlined in section 3.2.2.

For brevity, we only explore these aims for a single reference category. If our proposed sample size calculation approach is successful at keeping the multinomial sub-model specific shrinkage factors with the chosen reference category above the pre-specified threshold, then this would extend to repeating the process with the other reference categories. Our proposition in section 3.2.4 of repeating the process for each reference category, and taking the maximum sample size across all of them, will then be sufficient for targeting all the multinomial sub-model specific shrinkage factors to be above the pre-specified threshold.

As a brief overview, we simulated artificial development datasets for a variety of sample sizes, two of which met our proposed sample size criteria from section 3.2. We then fit multinomial logistic regression models and distinct binary logistic regression models to these datasets. Sub-model specific shrinkage factors ($S_{MN,k,r}$) and distinct logistic shrinkage factors ($S_{DL,k,r}$) were obtained using the validation process given in section 3.2.2, in a large simulated dataset representing the population of interest. This involved fitting the recalibration framework of van Hoorde et al.,[7] and binary logistic recalibration techniques[8–11]. $S_{MN,k,r}$ and $S_{DL,k,r}$ were then compared to the targeted threshold of 0.9, and each other. Sample sizes using the proposed criteria were calculated based on 'available information' from models which we fitted to another artificial dataset, to mimic the process which would happen in practice, of obtaining estimates of $R^2_{CS}$ from previously available models.

### 9.1.2. Data generating mechanisms and calculation of $N_{MN}$ and $N_{DL}$

Data were generated for a three-category outcome, $Y \in \{1,2,3\}$. Five independent predictor variables $X_1$, $X_2$, $X_3$, $X_4$, $X_5$ were generated by simulating $X_i \sim N(0,1), i \in \{1,2,3,4,5\}$, for N individuals (values of N described below). Ten different scenarios were considered by varying the coefficients of these predictor variables, with the aim of varying the outcome proportion in the different outcome categories. The model coefficients are denoted by $\beta_{0,2}$, $\beta_{1,2}$, $\beta_{2,2}$, $\beta_{3,2}$, $\beta_{4,2}$, $\beta_{5,2}$ and $\beta_{0,3}$, $\beta_{1,3}$, $\beta_{2,3}$, $\beta_{3,3}$, $\beta_{4,3}$, $\beta_{5,3}$, with values considered in the simulation scenarios given in Supplementary Table S1. For each individual, outcome data were simulated through a multinomial distribution with probabilities based on the following set of equations:[12]

$$P(Y=1) = \frac{1}{1+\exp\left(\beta_{0,2}+\sum_{q=1}^{5}\beta_{q,2}X_q\right)+\exp\left(\beta_{0,3}+\sum_{q=1}^{5}\beta_{q,3}X_q\right)}$$

$$P(Y=2) = \frac{\exp\left(\beta_{0,2}+\sum_{q=1}^{5}\beta_{q,2}X_q\right)}{1+\exp\left(\beta_{0,2}+\sum_{q=1}^{5}\beta_{q,2}X_q\right)+\exp\left(\beta_{0,3}+\sum_{q=1}^{5}\beta_{q,3}X_q\right)}$$

$$P(Y=3) = \frac{\exp\left(\beta_{0,3}+\sum_{q=1}^{5}\beta_{q,3}X_q\right)}{1+\exp\left(\beta_{0,2}+\sum_{q=1}^{5}\beta_{q,2}X_q\right)+\exp\left(\beta_{0,3}+\sum_{q=1}^{5}\beta_{q,3}X_q\right)}$$

The values of $\beta$ and corresponding category outcome proportions in each scenario are provided in Supplementary Table S1. Scenarios 7 to 12 have covariate effects twice as big as scenarios 1 to 6. The intercepts of each outcome category model were selected so that simulation scenarios 1 and 7 represent a balanced outcome proportion situation, simulation scenarios 2 and 8 represent a 'one lower' outcome proportion situation, simulation scenarios 3 and 9 represent an 'all different' outcome proportion situation, simulation scenarios 4 and 10 represent a 'one rare category' situation, and simulation scenarios 5 and 11 represent a 'one very rare category' situation. The outcome proportion of the 'very rare category' in scenarios 6 and 12 (6%), was chosen to match the outcome proportion of borderline malignant (the rarest category), in the case study of ovarian cancer by de Jong et al.[13] Without loss of generality, we chose $Y = 1$ to always be the most prevalent category, and this was always chosen to be the reference category when fitting models and calibrating. This is often a reasonable approach to take in clinical prediction, when the covariate effects are not of direct interest.

We denote $N_{MN}$ to be the minimum sample size calculated by applying criterion (i) using the approach from section3.2.1, and $N_{DL}$ using the approach from section 3.2.2 based on distinct logistic regression models. For both we take category 1 to be the reference category. For each simulation scenario, we generated 1000 development datasets of size $N = 100, 250, 500, 1000$, $N_{MN}$ and $N_{DL}$. For each scenario, a validation dataset of size 500,000 was generated under the same data generating mechanisms as described above to represent a population, in which the required shrinkage of the models was assessed.

<u>Calculation of $N_{MN}$ and $N_{DL}$</u>

To calculate the minimum required sample sizes ($N_{MN}$ and $N_{DL}$) to satisfy criterion (i) for each scenario, we mimicked what would happen in practice by basing the sample size calculation off available information that would be available a priori. Specifically, for each scenario a multinomial logistic regression model and a set of distinct logistic regression models were fitted to a cohort of 500,000 individuals. This dataset is different from the validation dataset. The $R^2_{CS_adj}$ of these models was then assumed to be "publically available" and was utilised in the sample size calculations for each scenario, following the processes outlined in sections 3.2.1 and 3.2.2. A cohort size of 500,000 was chosen to ensure the value of $R^2_{CS_adj}$ used in the sample size calculation was accurate. Both sample sizes were calculated with the intention targeting a shrinkage factor of 0.9. The exact process for calculating these is detailed in the supplementary methods.

<u>For each scenario, $N_{MN}$ was calculated as follows:</u>

1. Generate a cohort of size 500,000 using the aforementioned data generating mechanisms

2. Fit a multinomial model with $X_1$ to $X_5$ as predictors (equation (6)), and fit an intercept only multinomial model (null model) to this dataset.

3. Calculate the likelihood ratio as $LR = -2(loglikelihood_{null} - loglikelihood_{full})$

4. Calculate $R^2_{CS_app} = 1 - \exp(LR/500000)$

5. Calculate $S_{VH_MN} = 1 + \frac{10}{500000 * \log(1 - R^2_{CS_app})}$

6. Calculate $R^2_{CS_adj} = S_{VH_MN} \times R^2_{CS_app}$

7. Calculate $N_{MN} = \frac{10}{(0.9-1)*log(1-R^2_{CS_adj}/0.9)}$

Note, that $p$ = 10, as although there are only 5 predictors, there are 10 predictor parameters to be calculated in the multinomial model. Step 7 could be replaced by using the *pmsampsize* package in R,[14] using the value of $R^2_{adj}$ calculated from step 6.

<u>For each scenario, $N_{DL}$ was calculated as follows:</u>

1. Generate a cohort of size 500000 using the aforementioned data generating mechanisms

2. Fit distinct logistic regression models with $X_1$ to $X_5$ as predictors (full models), and intercept only distinct logistic regression models (null models), to predict $\log\left(\frac{P(Y=2)}{P(Y=1)}\right)$ and $\log\left(\frac{P(Y=3)}{P(Y=1)}\right)$, on the subsets of the cohort which have the appropriate outcomes. These models match the sub-models from the multinomial framework.

3. Calculate the corresponding likelihood ratio from the models in step 2, as $LR_{k,1} = -2(loglikelihood_{null,k,1} - loglikelihood_{full,k,1})$, for each model $k$.

Repeat the following steps 4 to 7 for both models ($k = 1,2$) to calculate $n_{2,1}$ and $n_{3,1}$ from Box 1 (summary box):

4. Calculate $R^2_{CS_app,k,1} = 1 - \exp\left(LR_{k,1}/\sum_{i=1}^{500000} I(Y \in \{1,k\})\right)$

5. Calculate $S_{VH_DL,k,1} = 1 + \frac{5}{\sum_{i=1}^{500000} I(Y\in\{1,k\})*\log(1-R^2_{CS_app,k})}$

6. Calculate $R^2_{CS_adj,k,1} = S_{VH_DL,k,1} \times R^2_{CS_app,k,1}$

7. Calculate $\omega_{k,i} = \frac{\sum_{i=1}^{500000} I(Y\in\{1,k\})}{500000}$

8. Calculate $m_{k,1} = \frac{5}{(0.9-1)*\log(1-R^2_{CS_adj,k,1}/0.9)}$

9. Calculate $n_{k,1} = m_{k,1}/\omega_{k,1}$

8. Take $N_{DL} = \max(n_{2,1}, n_{3,1})$.

Note that $p$ = 5 in these calculations, as there are five predictors in each model, which is independent from the other model. In step 7, the sample size required to target the threshold in each distinct logistic regression model, is divided by the proportion of individuals from the initial cohort that would be used in this sub-model.

### 9.1.3. Estimands

The Estimands of interest were:

$S_{MN,2,1}$ and $S_{MN,3,1}$: sub-model specific shrinkage factors. These values therefore represented the level of shrinkage required when the multinomial model was implemented in the population of interest.

$S_{DL,2,1}$ and $S_{DL,3,1}$: distinct logistic shrinkage factors

$S_{VH_MN}$: heuristic shrinkage factor of the multinomial model

$S_{VH_DL,2,1}$ and $S_{VH_DL,3,1}$: heuristic shrinkage factors of the distinct logistic regression models

### 9.1.4. Methods and models for comparison

A multinomial logistic regression and distinct logistic regression models with $X_1$ to $X_5$ as predictors were fitted to each development dataset. The sub-model specific shrinkage factors of the multinomial models ($S_{MN,2,1}$ and $S_{MN,3,1}$) were calculated by implementing the calibration framework of van Hoorde et al.[7] in the validation cohort, with category 1 as the reference category. The distinct logistic shrinkage factors ($S_{DL,2,1}$ and $S_{DL,3,1}$), were calculated using binary logistic calibration techniques[8–11] to calibrate the models in the validation cohort. This follows the process outlined in section 3.2.2. $S_{VH_MN}$ was the heuristic shrinkage factor of the multinomial model, and $S_{VH_DL,2,1}$ and $S_{VH_DL,3,1}$ were the heuristic shrinkage factors of the distinct logistic regression models, calculated using equations (13) and (3) respectively.

The methods used to produce the estimands (described in this section) were not being compared in this simulation. Instead, we were comparing the value of these estimands when different sample sizes were used to generate the dataset. In particular, when $N = N_{MN}$ or $N_{DL}$.

### 9.1.5. Performance measures and comparisons

We do not use standard performance measures given we are not comparing the ability of different methods to estimate an estimand of interest. The main performance measures are the mean and median of the sub-model specific shrinkage factors and the distinct logistic shrinkage factors. We compare these estimates with the targeted threshold, and each other, in three main ways listed here:

1) Comparison of the median and mean (across the 1000 simulations) of the sub-model specific shrinkage factors, $S_{MN,2,1}$ and $S_{MN,3,1}$, to the threshold 0.9. This assess the ultimate aim of sample size criteria (i) for the multinomial model, which is to target $S_{MN,2,1}$ and $S_{MN,3,1}$ to be at 0.9 on average.

2) Comparison of the sub-model specific shrinkage factors of the multinomial model ($S_{MN,2,1}$ and $S_{MN,3,1}$) and distinct logistic shrinkage factors ($S_{DL,21}$ and $S_{DL,3,1}$). The process for deriving $N_{DL}$ given in section 3.2.2 was based on a proposition that the sub-model specific shrinkage factors of the multinomial model will tend towards the distinct logistic shrinkage factors as $N \to \infty$. This comparison will assess this proposition.

3) Comparison of the distinct logistic shrinkage factors ($S_{DL,2,1}$ and $S_{DL,3,1}$), and the corresponding heuristic shrinkage factors ($S_{VH_DL,2,1}$ and $S_{VH_DL,3,1}$). Given that the sample size criterion are based on ensuring the heuristic shrinkage factors are close to the pre-specified value, but the ultimate aim is ensure the actual required shrinkage is close to the pre-specified value, it was important to evaluate the agreement between the two.

Note that the variance of the sub-model specific shrinkage factors and distinct logistic shrinkage factors are not relevant to our research question, because the proposed sample size criteria do not target anything about the variance of the shrinkage factors. However, it is of wider interest to researchers to know how variable the level of required shrinkage in practice is. We therefore present data on the variability of $S_{MN,2,1}$ and $S_{MN,3,1}$ in Appendix section 2.

## 9.2. Simulation results and discussion

In this section we are interested in comparing the median/mean of the shrinkage factors, and whenever we refer to median/mean, this is the median/mean of the stated values (i.e. $S_{MN,2,1}$ and $S_{MN,3,1}$) across the 1000 simulations.

### 9.2.1. Comparison of the median and mean of the sub-model specific shrinkage factors, $S_{MN,2,1}$ and $S_{MN,3,1}$, to the threshold 0.9.

Our results suggest that basing criterion (i) on distinct logistic regressions ($N = N_{DL}$) achieved the aim of ensuring that the median/mean of the sub-model specific shrinkage factors of the multinomial model were $\geq 0.9$. In contrast, and as expected, applying the sample size formula directly to the multinomial model ($N = N_{MN}$) only achieved this aim in some scenarios.

For example, in scenarios 1 to 3 the values of $N_{MN}$ and $N_{DL}$ were very similar, and when $N = N_{MN}$ or $N_{DL}$ the median of $S_{MN,2,1}$ and $S_{MN,3,1}$ were both close to 0.9 (Supplementary Table S2). However, in scenarios 4 and 5 we observed larger differences between $N_{MN}$ and $N_{DL}$. When $N = N_{MN}$, median $S_{MN,2,1}$ was notably larger than 0.9, whereas median $S_{MN,3,1}$ was notably lower than 0.9. Importantly, this means the sample size criterion was not achieved in every sub-model. In contrast, when $N = N_{DL}$ the median $S_{MN,2,1}$ and $S_{MN,3,1}$ were both $\geq 0.9$, meaning the sample size criterion was achieved in every sub-model. Furthermore, the model with the smaller sub-model specific shrinkage factor ($S_{MN,3,1}$) was very close to 0.9, indicating the number of individuals included was about as small as it could be to still ensure criterion (i) is met.

The targeted level of shrinkage ($S_{MN,2,1}$ and $S_{MN,3,1}$ both at or above $0.9$) was consistently not met for scenarios 7 to 12 (Supplementary Table S3) for $N = N_{MN}$ or $N_{DL}$. This finding appeared to be due to poor agreement between the distinct logistic shrinkage factors and the corresponding heuristic shrinkage factor in these scenarios, and is discussed in more detail in section 9.2.3.

### 9.2.2. Comparison of the sub-model specific shrinkage factors of the multinomial model ($S_{MN,2,1}$ and $S_{MN,3,1}$) and shrinkage factors of the distinct logistic models ($S_{DL,2,1}$ and $S_{DL,3,1}$).

Supplementary Table S2 shows there was strong agreement between the sub-model specific shrinkage factors of the multinomial model ($S_{MN,2,1}$ and $S_{MN,3,1}$), and the distinct logistic shrinkage factors ($S_{DL,2,1}$ and $S_{DL,3,1}$), with closer agreement as $N$ increased (as expected, since the two are identical as $N \to \infty$). Importantly, for $N = N_{DL}$ there appeared to be good agreement between the two, meaning the asymptotic equivalence was holding when $N = N_{DL}$. This supports the reasoning behind targeting the shrinkage of distinct logistic regression models, when the underlying aim is to target the multinomial sub-model specific shrinkage factors.

### 9.2.3. Comparison of the distinct logistic shrinkage factors ($S_{DL,2,1}$ and $S_{DL,3,1}$), and the corresponding heuristic shrinkage factors ($S_{VH_DL,2,1}$ and $S_{VH_DL,3,1}$).

For scenarios 1 to 6 there was good agreement between the medians of $S_{DL,2,1}$ and $S_{DL,3,1}$ with the medians of $S_{VH_DL,2,1}$ and $S_{VH_DL,3,1}$, with no consistent pattern in the direction of the disagreement (Supplementary Table S2). There were slightly bigger differences in the mean, which is likely because $S_{VH_DL,2,1}$ and $S_{VH_DL,3,1}$ are bounded at 1, resulting in a skewed distribution, which is not the case for $S_{DL,2,1}$ and $S_{DL,3,1}$.

On the contrary, there was relatively poor agreement in the average $S_{DL,2,1}$ and $S_{DL,3,1}$ with the average $S_{VH_DL,2,1}$ and $S_{VH_DL,3,1}$ in scenarios 7 to 12 (Supplementary Table S3). Specifically, the medians and means of $S_{VH_DL,2,1}$ and $S_{VH_DL,3,1}$ were consistently higher than those of $S_{DL,2,1}$ and $S_{DL,3,1}$. These results suggest that our proposed criterion successfully ensured the median/mean value of both heuristic shrinkage factors was $\geq 0.9$, but it did not ensure the median/mean value of the shrinkage required in the validation cohort was $\geq 0.9$. This is an important because it is not unique to multinomial models. The major difference between scenarios 1 to 6 (where there was good agreement) and 7 to 12 (where $S_{VH_DL,k,r}$ was consistently higher than $S_{DL,k,r}$) was that the covariate effect sizes were doubled in the latter. This in turn means the required sample sizes for scenarios 7 to 12 were much smaller than their counterparts in scenarios 1 to 6, as larger covariate effects require lower sample sizes to estimate. This simulation indicates that with these larger effect sizes and smaller sample sizes, the heuristic shrinkage factor no longer estimates the shrinkage factor upon validation unbiasedly. It may therefore not be enough to target the average heuristic shrinkage factor to be at $0.9$ in scenarios where covariate effects are expected a priori to be large. The mechanism behind this finding is unclear, and further work is required to understand this observation.

## 9.3. Overall conclusions

1) We suggest basing the sample size for criterion (i) on the shrinkage of distinct logistic regression models (sections **3.2.2** - **3.2.4**). In our simulations, $N_{DL}$ ensured the average value of all the sub-model specific shrinkage factors was $\geq 0.9$, as targeted. We advise against basing the sample size on the overall shrinkage of the multinomial model (section **3.2.1**). In our simulations, $N = N_{MN}$ did not always result in the desired level of shrinkage in all sub-models. There were no scenarios that we considered where $N = N_{MN}$ outperformed $N = N_{DL}$ with regards to this aim of the sample size criterion.

2) There was good agreement between sub-model specific shrinkage factors ($S_{MN,2,1}$ and $S_{MN,3,1}$), and the distinct logistic shrinkage factors ($S_{DL,2,1}$ and $S_{DL,3,1}$), supporting the claim that these are equivalent as $N \to \infty$.

3) There was a poor agreement between the heuristic shrinkage factors and the distinct logistic shrinkage factors at model validation in scenarios 7 – 12 (increased covariate effect sizes), which resulted in sub-model specific shrinkage factors that were below the targeted threshold. This will need to be explored in future work and is relevant to binary logistic regression too.

## 9.4. Tables

*Supplementary Table S1: Beta coefficients and resulting outcome proportions*

| | | Scenario | | | | | | | | | | | |
|---|---|---|---|---|---|---|---|---|---|---|---|---|---|
| | | Smaller covariate effects | | | | | | Larger covariate effects | | | | | |
| | | 1 | 2 | 3 | 4 | 5 | 6 | 7 | 8 | 9 | 10 | 11 | 12 |
| Coefficients | $\beta_{0,2}$ | 0 | 0 | -0.35 | -0.4 | -0.53 | -2.9 | 0 | 0 | -0.4 | -0.5 | -0.65 | -3.5 |
| | $\beta_{1,2}$ | 0.5 | 0.5 | 0.5 | 0.5 | 0.5 | 0.5 | 1 | 1 | 1 | 1 | 1 | 1 |
| | $\beta_{2,2}$ | -0.25 | -0.25 | -0.25 | -0.25 | -0.25 | -0.25 | -0.5 | -0.5 | -0.5 | -0.5 | -0.5 | -0.5 |
| | $\beta_{3,2}$ | -0.125 | -0.125 | -0.125 | -0.125 | -0.125 | -0.125 | -0.25 | -0.25 | -0.25 | -0.25 | -0.25 | -0.25 |
| | $\beta_{4,2}$ | 0.25 | 0.25 | 0.25 | 0.25 | 0.25 | 0.25 | 0.5 | 0.5 | 0.5 | 0.5 | 0.5 | 0.5 |
| | $\beta_{5,2}$ | 0.375 | 0.375 | 0.375 | 0.375 | 0.375 | 0.375 | 0.75 | 0.75 | 0.75 | 0.75 | 0.75 | 0.75 |
| | $\beta_{0,3}$ | 0 | -0.75 | -0.85 | -1.7 | -2.4 | -2.9 | -1 | -1 | -1 | -2 | -2.85 | -3.5 |
| | $\beta_{1,3}$ | 0.375 | 0.375 | 0.375 | 0.375 | 0.375 | 0.375 | 0.75 | 0.75 | 0.75 | 0.75 | 0.75 | 0.75 |
| | $\beta_{2,3}$ | -0.5 | -0.5 | -0.5 | -0.5 | -0.5 | -0.5 | -1 | -1 | -1 | -1 | -1 | -1 |
| | $\beta_{3,3}$ | -0.25 | -0.25 | -0.25 | -0.25 | -0.25 | -0.25 | -0.5 | -0.5 | -0.5 | -0.5 | -0.5 | -0.5 |
| | $\beta_{4,3}$ | -0.375 | -0.375 | -0.375 | -0.375 | -0.375 | -0.375 | -0.75 | -0.75 | -0.75 | -0.75 | -0.75 | -0.75 |
| | $\beta_{5,3}$ | 0.125 | 0.125 | 0.125 | 0.125 | 0.125 | 0.125 | 0.25 | 0.25 | 0.25 | 0.25 | 0.25 | 0.25 |
| Outcome frequencies | Y=1 | 33% | 40% | 45% | 52% | 58% | 88% | 33% | 40% | 45% | 52% | 58% | 88% |
| | Y=2 | 33% | 40% | 33% | 36% | 36% | 6% | 33% | 40% | 33% | 36% | 36% | 6% |
| | Y=3 | 33% | 19% | 21% | 11% | 6% | 6% | 34% | 19% | 21% | 11% | 6% | 6% |

*Supplementary Table S2: Median (mean) sub-model specific shrinkage factors and distinct logistic shrinkage factors, and heuristic shrinkage factors from multinomial and distinct logistic regression models, scenarios 7 to 12*

| Scenario 1 (all same) | | | | | | | |
|---|---|---|---|---|---|---|---|
| | Multinomial | | | Distinct logistic | | | |
| N | $S_{MN,2,1}$ | $S_{MN,3,1}$ | $S_{VH_MN}$ | $S_{DL,2,1}$ | $S_{DL,3,1}$ | $S_{VH_DL,2,1}$ | $S_{VH_DL,3,1}$ |
| 250 | 0.783 (0.807) | 0.806 (0.828) | 0.802 (0.791) | 0.779 (0.806) | 0.785 (0.821) | 0.785 (0.759) | 0.811 (0.784) |
| 500 | 0.886 (0.904) | 0.892 (0.905) | 0.891 (0.888) | 0.884 (0.904) | 0.887 (0.904) | 0.882 (0.874) | 0.895 (0.889) |
| 1000 | 0.943 (0.948) | 0.943 (0.95) | 0.942 (0.942) | 0.933 (0.944) | 0.937 (0.95) | 0.938 (0.936) | 0.945 (0.944) |
| $N_{MN}$ (541) | 0.893 (0.905) | 0.899 (0.912) | 0.899 (0.896) | 0.879 (0.899) | 0.899 (0.908) | 0.892 (0.884) | 0.902 (0.898) |
| $N_{DL}$ (576) | 0.906 (0.92) | 0.902 (0.919) | 0.903 (0.901) | 0.903 (0.916) | 0.898 (0.918) | 0.895 (0.889) | 0.909 (0.903) |
| Scenario 2 (one lower) | | | | | | | |
| | Multinomial | | | Distinct logistic | | | |
| N | $S_{MN,2,1}$ | $S_{MN,3,1}$ | $S_{VH_MN}$ | $S_{DL,2,1}$ | $S_{DL,3,1}$ | $S_{VH_DL,2,1}$ | $S_{VH_DL,3,1}$ |
| 250 | 0.811 (0.829) | 0.771 (0.8) | 0.793 (0.78) | 0.814 (0.841) | 0.763 (0.794) | 0.811 (0.793) | 0.768 (0.736) |
| 500 | 0.908 (0.92) | 0.887 (0.899) | 0.884 (0.881) | 0.91 (0.929) | 0.88 (0.894) | 0.899 (0.893) | 0.871 (0.864) |
| 1000 | 0.951 (0.957) | 0.934 (0.946) | 0.94 (0.939) | 0.953 (0.963) | 0.931 (0.945) | 0.948 (0.946) | 0.934 (0.931) |
| $N_{MN}$ (569) | 0.914 (0.921) | 0.891 (0.909) | 0.899 (0.896) | 0.917 (0.931) | 0.883 (0.908) | 0.911 (0.907) | 0.889 (0.879) |
| $N_{DL}$ (628) | 0.926 (0.936) | 0.908 (0.923) | 0.906 (0.904) | 0.935 (0.944) | 0.909 (0.919) | 0.918 (0.915) | 0.894 (0.889) |
| Scenario 3 (all different) | | | | | | | |
| | Multinomial | | | Distinct logistic | | | |
| N | $S_{MN,2,1}$ | $S_{MN,3,1}$ | $S_{VH_MN}$ | $S_{DL,2,1}$ | $S_{DL,3,1}$ | $S_{VH_DL,2,1}$ | $S_{VH_DL,3,1}$ |
| 250 | 0.801 (0.826) | 0.772 (0.802) | 0.795 (0.781) | 0.809 (0.834) | 0.767 (0.8) | 0.808 (0.787) | 0.787 (0.756) |
| 500 | 0.899 (0.912) | 0.887 (0.901) | 0.884 (0.881) | 0.911 (0.921) | 0.879 (0.896) | 0.894 (0.889) | 0.88 (0.872) |
| 1000 | 0.939 (0.953) | 0.941 (0.948) | 0.939 (0.939) | 0.946 (0.959) | 0.936 (0.942) | 0.946 (0.944) | 0.937 (0.936) |
| $N_{MN}$ (566) | 0.921 (0.925) | 0.89 (0.906) | 0.897 (0.894) | 0.919 (0.929) | 0.882 (0.897) | 0.906 (0.902) | 0.895 (0.888) |
| $N_{DL}$ (582) | 0.911 (0.93) | 0.896 (0.909) | 0.9 (0.897) | 0.919 (0.937) | 0.887 (0.902) | 0.908 (0.903) | 0.898 (0.891) |
| Scenario 4 (one rare category) | | | | | | | |
| | Multinomial | | | Distinct logistic | | | |
| N | $S_{MN,2,1}$ | $S_{MN,3,1}$ | $S_{VH_MN}$ | $S_{DL,2,1}$ | $S_{DL,3,1}$ | $S_{VH_DL,2,1}$ | $S_{VH_DL,3,1}$ |
| 250 | 0.811 (0.837) | 0.699 (0.726) | 0.773 (0.76) | 0.825 (0.853) | 0.679 (0.713) | 0.827 (0.808) | 0.699 (0.638) |
| 500 | 0.899 (0.914) | 0.83 (0.851) | 0.873 (0.869) | 0.91 (0.925) | 0.83 (0.845) | 0.907 (0.901) | 0.825 (0.809) |
| 1000 | 0.952 (0.958) | 0.916 (0.931) | 0.932 (0.931) | 0.957 (0.963) | 0.909 (0.925) | 0.951 (0.95) | 0.906 (0.899) |
| $N_{MN}$ (648) | 0.933 (0.943) | 0.865 (0.885) | 0.899 (0.896) | 0.939 (0.952) | 0.858 (0.883) | 0.925 (0.921) | 0.862 (0.85) |
| $N_{DL}$ (901) | 0.936 (0.95) | 0.888 (0.909) | 0.927 (0.925) | 0.943 (0.958) | 0.886 (0.905) | 0.946 (0.945) | 0.9 (0.893) |
| Scenario 5 (one very rare category) | | | | | | | |
| | Multinomial | | | Distinct logistic | | | |
| N | $S_{MN,2,1}$ | $S_{MN,3,1}$ | $S_{VH_MN}$ | $S_{DL,2,1}$ | $S_{DL,3,1}$ | $S_{VH_DL,2,1}$ | $S_{VH_DL,3,1}$ |
| 250 | 0.8 (0.829) | 0.594 (0.622) | 0.76 (0.743) | 0.82 (0.844) | 0.583 (0.612) | 0.833 (0.815) | 0.598 (0.498) |
| 500 | 0.909 (0.915) | 0.772 (0.799) | 0.861 (0.856) | 0.92 (0.925) | 0.757 (0.793) | 0.906 (0.903) | 0.743 (0.695) |
| 1000 | 0.948 (0.955) | 0.878 (0.901) | 0.926 (0.925) | 0.952 (0.96) | 0.873 (0.897) | 0.952 (0.951) | 0.859 (0.845) |
| $N_{MN}$ (706) | 0.93 (0.943) | 0.846 (0.868) | 0.898 (0.895) | 0.94 (0.952) | 0.843 (0.867) | 0.933 (0.93) | 0.808 (0.78) |
| $N_{DL}$ (1458) | 0.967 (0.97) | 0.913 (0.931) | 0.948 (0.948) | 0.971 (0.974) | 0.918 (0.93) | 0.967 (0.966) | 0.899 (0.893) |
| Scenario 6 (two very rare categories) | | | | | | | |
| | Multinomial | | | Distinct logistic | | | |
| N | $S_{MN,2,1}$ | $S_{MN,3,1}$ | $S_{VH_MN}$ | $S_{DL,2,1}$ | $S_{DL,3,1}$ | $S_{VH_DL,2,1}$ | $S_{VH_DL,3,1}$ |
| 250 | 0.58 (0.61) | 0.585 (0.609) | 0.582 (0.529) | 0.589 (0.618) | 0.592 (0.617) | 0.537 (0.356) | 0.602 (0.489) |
| 500 | 0.764 (0.791) | 0.744 (0.78) | 0.738 (0.719) | 0.77 (0.798) | 0.752 (0.788) | 0.715 (0.66) | 0.762 (0.71) |
| 1000 | 0.861 (0.889) | 0.857 (0.881) | 0.855 (0.847) | 0.866 (0.893) | 0.86 (0.885) | 0.846 (0.831) | 0.864 (0.85) |
| $N_{MN}$ (1558) | 0.924 (0.94) | 0.902 (0.92) | 0.9 (0.898) | 0.924 (0.943) | 0.909 (0.925) | 0.895 (0.888) | 0.909 (0.903) |
| $N_{DL}$ (1616) | 0.927 (0.943) | 0.911 (0.924) | 0.902 (0.9) | 0.931 (0.946) | 0.917 (0.928) | 0.897 (0.891) | 0.91 (0.905) |

$S_{MN,k,r}$, multinomial sub-model specific shrinkage factors; $S_{VH_MN}$, heuristic shrinkage factor of multinomial model; $S_{DL,k,r}$, distinct logistic shrinkage factors; $S_{VH_DL,k,r}$, heuristic shrinkage factor of distinct logistic models.

*Supplementary Table S3: Median (mean) sub-model specific shrinkage factors and distinct logistic shrinkage factors, and heuristic shrinkage factors from multinomial and distinct logistic regression models, scenarios 7 to 12*

| Scenario 7 (all same) | | | | | | | |
|---|---|---|---|---|---|---|---|
| | Multinomial | | | Distinct logistic | | | |
| N | $S_{MN,2,1}$ | $S_{MN,3,1}$ | $S_{VH_MN}$ | $S_{DL,2,1}$ | $S_{DL,3,1}$ | $S_{VH_DL,2,1}$ | $S_{VH_DL,3,1}$ |
| 250 | 0.894 (0.904) | 0.898 (0.906) | 0.924 (0.923) | 0.879 (0.891) | 0.883 (0.894) | 0.918 (0.914) | 0.925 (0.922) |
| 500 | 0.942 (0.947) | 0.936 (0.944) | 0.961 (0.961) | 0.934 (0.939) | 0.923 (0.932) | 0.957 (0.956) | 0.962 (0.961) |
| 1000 | 0.97 (0.973) | 0.964 (0.967) | 0.98 (0.98) | 0.958 (0.965) | 0.958 (0.963) | 0.978 (0.978) | 0.98 (0.98) |
| $N_{MN}$ (177) | 0.846 (0.862) | 0.849 (0.861) | 0.897 (0.895) | 0.831 (0.844) | 0.826 (0.841) | 0.887 (0.882) | 0.899 (0.894) |
| $N_{DL}$ (196) | 0.866 (0.886) | 0.874 (0.884) | 0.906 (0.902) | 0.856 (0.875) | 0.846 (0.867) | 0.895 (0.889) | 0.906 (0.902) |
| Scenario 8 (one lower) | | | | | | | |
| | Multinomial | | | Distinct logistic | | | |
| N | $S_{MN,2,1}$ | $S_{MN,3,1}$ | $S_{VH_MN}$ | $S_{DL,2,1}$ | $S_{DL,3,1}$ | $S_{VH_DL,2,1}$ | $S_{VH_DL,3,1}$ |
| 250 | 0.895 (0.904) | 0.874 (0.891) | 0.92 (0.919) | 0.897 (0.905) | 0.85 (0.865) | 0.93 (0.928) | 0.909 (0.905) |
| 500 | 0.941 (0.95) | 0.932 (0.94) | 0.959 (0.958) | 0.945 (0.948) | 0.917 (0.925) | 0.964 (0.963) | 0.953 (0.952) |
| 1000 | 0.968 (0.971) | 0.962 (0.965) | 0.979 (0.979) | 0.972 (0.97) | 0.96 (0.961) | 0.982 (0.982) | 0.976 (0.975) |
| $N_{MN}$ (189) | 0.866 (0.88) | 0.836 (0.851) | 0.897 (0.895) | 0.862 (0.879) | 0.812 (0.832) | 0.909 (0.905) | 0.884 (0.876) |
| $N_{DL}$ (219) | 0.883 (0.892) | 0.874 (0.881) | 0.91 (0.908) | 0.885 (0.893) | 0.84 (0.858) | 0.921 (0.918) | 0.896 (0.891) |
| Scenario 9 (all different) | | | | | | | |
| | Multinomial | | | Distinct logistic | | | |
| N | $S_{MN,2,1}$ | $S_{MN,3,1}$ | $S_{VH_MN}$ | $S_{DL,2,1}$ | $S_{DL,3,1}$ | $S_{VH_DL,2,1}$ | $S_{VH_DL,3,1}$ |
| 250 | 0.895 (0.906) | 0.888 (0.896) | 0.921 (0.92) | 0.888 (0.902) | 0.861 (0.878) | 0.927 (0.925) | 0.918 (0.914) |
| 500 | 0.952 (0.957) | 0.943 (0.953) | 0.959 (0.958) | 0.953 (0.957) | 0.934 (0.944) | 0.962 (0.961) | 0.957 (0.956) |
| 1000 | 0.973 (0.977) | 0.968 (0.97) | 0.979 (0.979) | 0.976 (0.979) | 0.962 (0.965) | 0.981 (0.981) | 0.978 (0.978) |
| $N_{MN}$ (186) | 0.86 (0.873) | 0.849 (0.863) | 0.898 (0.895) | 0.861 (0.87) | 0.831 (0.846) | 0.905 (0.9) | 0.893 (0.886) |
| $N_{DL}$ (198) | 0.86 (0.877) | 0.856 (0.873) | 0.904 (0.901) | 0.857 (0.873) | 0.839 (0.856) | 0.911 (0.906) | 0.898 (0.892) |
| Scenario 10 (one rare category) | | | | | | | |
| | Multinomial | | | Distinct logistic | | | |
| N | $S_{MN,2,1}$ | $S_{MN,3,1}$ | $S_{VH_MN}$ | $S_{DL,2,1}$ | $S_{DL,3,1}$ | $S_{VH_DL,2,1}$ | $S_{VH_DL,3,1}$ |
| 250 | 0.899 (0.904) | 0.855 (0.874) | 0.913 (0.911) | 0.904 (0.912) | 0.834 (0.849) | 0.934 (0.932) | 0.884 (0.875) |
| 500 | 0.949 (0.953) | 0.924 (0.935) | 0.954 (0.954) | 0.953 (0.956) | 0.924 (0.933) | 0.966 (0.965) | 0.938 (0.936) |
| 1000 | 0.973 (0.976) | 0.958 (0.965) | 0.977 (0.977) | 0.978 (0.98) | 0.954 (0.959) | 0.983 (0.982) | 0.968 (0.968) |
| $N_{MN}$ (209) | 0.876 (0.886) | 0.824 (0.842) | 0.898 (0.895) | 0.883 (0.892) | 0.792 (0.819) | 0.922 (0.919) | 0.864 (0.851) |
| $N_{DL}$ (289) | 0.908 (0.922) | 0.865 (0.886) | 0.924 (0.922) | 0.913 (0.926) | 0.859 (0.87) | 0.942 (0.94) | 0.897 (0.89) |
| Scenario 11 (one very rare category) | | | | | | | |
| | Multinomial | | | Distinct logistic | | | |
| N | $S_{MN,2,1}$ | $S_{MN,3,1}$ | $S_{VH_MN}$ | $S_{DL,2,1}$ | $S_{DL,3,1}$ | $S_{VH_DL,2,1}$ | $S_{VH_DL,3,1}$ |
| 250 | 0.897 (0.907) | 0.807 (0.826) | 0.903 (0.901) | 0.908 (0.918) | 0.773 (0.793) | 0.935 (0.933) | 0.827 (0.805) |
| 500 | 0.952 (0.957) | 0.903 (0.916) | 0.949 (0.949) | 0.957 (0.962) | 0.879 (0.895) | 0.966 (0.966) | 0.907 (0.901) |
| 1000 | 0.977 (0.978) | 0.955 (0.964) | 0.974 (0.974) | 0.976 (0.98) | 0.947 (0.953) | 0.983 (0.983) | 0.951 (0.95) |
| $N_{MN}$ (233) | 0.888 (0.902) | 0.791 (0.805) | 0.899 (0.896) | 0.902 (0.914) | 0.753 (0.772) | 0.931 (0.929) | 0.82 (0.794) |
| $N_{DL}$ (455) | 0.939 (0.944) | 0.887 (0.9) | 0.945 (0.944) | 0.942 (0.949) | 0.866 (0.877) | 0.964 (0.963) | 0.901 (0.893) |
| Scenario 12 (two very rare categories) | | | | | | | |
| | Multinomial | | | Distinct logistic | | | |
| N | $S_{MN,2,1}$ | $S_{MN,3,1}$ | $S_{VH_MN}$ | $S_{DL,2,1}$ | $S_{DL,3,1}$ | $S_{VH_DL,2,1}$ | $S_{VH_DL,3,1}$ |
| 250 | 0.779 (0.789) | 0.787 (0.806) | 0.826 (0.816) | 0.775 (0.79) | 0.787 (0.811) | 0.813 (0.782) | 0.842 (0.817) |
| 500 | 0.871 (0.888) | 0.881 (0.896) | 0.905 (0.902) | 0.878 (0.888) | 0.884 (0.899) | 0.899 (0.891) | 0.913 (0.908) |
| 1000 | 0.926 (0.933) | 0.939 (0.942) | 0.951 (0.95) | 0.921 (0.933) | 0.941 (0.943) | 0.947 (0.945) | 0.955 (0.954) |
| $N_{MN}$ (470) | 0.862 (0.873) | 0.876 (0.89) | 0.9 (0.896) | 0.859 (0.873) | 0.878 (0.891) | 0.891 (0.883) | 0.908 (0.902) |
| $N_{DL}$ (505) | 0.875 (0.89) | 0.892 (0.903) | 0.906 (0.903) | 0.874 (0.891) | 0.893 (0.902) | 0.899 (0.891) | 0.915 (0.909) |

$S_{MN,k,r}$, multinomial sub-model specific shrinkage factors; $S_{VH_MN}$, heuristic shrinkage factor of multinomial model; $S_{DL,k,r}$, distinct logistic shrinkage factors; $S_{VH_DL,k,r}$, heuristic shrinkage factor of distinct logistic models.

# 10. Discussion of variability of the sub-model specific shrinkage factors

Supplementary Table S4 provides the full distribution of $S_{MN,2,1}$ and $S_{MN,3,1}$. We see that there is a reasonable amount of variability in the sub-model specific shrinkage factors across the 1000 simulations. When $N = N_{DL}$, the 25$^{th}$ percentile for the smaller sub-model specific shrinkage factor was approximately $0.825$, meaning in our simulation 25% of models meeting the sample size criteria, the shrinkage factor of that sub-model was $\leq 0.825$. This is an important result given that in practice you only get to recruit one cohort of individuals and build one model, and for many shrinkage methods the estimated shrinkage is inversely proportional to the actual amount of shrinkage required.[16,17] While important, it was also to be expected. The sample size criteria target the average shrinkage factor to be at a pre-specified threshold, in which they are successful. Our findings agree with previous work which reported on the variability in model performance when the sample size criteria are met.[18] We report on this to re-iterate to model users and developers that meeting the sample size criteria will not ensure the required shrinkage of your model is above $0.9$. Shrinkage techniques are therefore still advised when sample size criteria are met,[18] however even this is not a full-proof solution to the issue of overfitting.[16,19] If it is possible to use a sample size corresponding a higher threshold than $0.9$ this is encouraged, as it will reduce both the average and variability in the amount of shrinkage required. If feasible in the data collection process, an adaptive sample size approach[20] could be taken where the level of overfitting in the developed model is monitored, with recruitment/data collection stopping after the level of overfitting reaches $0.9$. This means the desired level of shrinkage should be achieved in the model of interest, which may not be the case when calculating a fixed sample size prior to model development.

*Supplementary Table S4: 2.5th, 25th, 50th, 75th, 97.5th percentile of sub-model specific shrinkage factors, $S_{MN,2,1}$ and $S_{MN,3,1}$, across the 1000 simulations, scenarios 1 to 6*

| Scenario 1 (all same) | | | | | | | | | | |
|---|---|---|---|---|---|---|---|---|---|---|
| | $S_{MN,2,1}$ | | | | | $S_{MN,3,1}$ | | | | |
| N | 2.5th | 25th | 50th | 75th | 97.5th | 2.5th | 25th | 50th | 75th | 97.5th |
| 250 | 0.549 | 0.693 | 0.783 | 0.907 | 1.193 | 0.571 | 0.711 | 0.806 | 0.914 | 1.230 |
| 500 | 0.684 | 0.815 | 0.886 | 0.982 | 1.211 | 0.695 | 0.814 | 0.892 | 0.978 | 1.199 |
| 1000 | 0.780 | 0.876 | 0.943 | 1.010 | 1.139 | 0.788 | 0.880 | 0.943 | 1.015 | 1.154 |
| $N_{MN}$ (541) | 0.687 | 0.810 | 0.893 | 0.980 | 1.210 | 0.702 | 0.830 | 0.899 | 0.987 | 1.179 |
| $N_{DL}$ (576) | 0.711 | 0.836 | 0.906 | 0.989 | 1.209 | 0.720 | 0.827 | 0.902 | 0.992 | 1.193 |
| Scenario 2 (one lower) | | | | | | | | | | |
| | $S_{MN,2,1}$ | | | | | $S_{MN,3,1}$ | | | | |
| N | 2.5th | 25th | 50th | 75th | 97.5th | 2.5th | 25th | 50th | 75th | 97.5th |
| 250 | 0.567 | 0.709 | 0.811 | 0.916 | 1.215 | 0.529 | 0.675 | 0.771 | 0.893 | 1.218 |
| 500 | 0.704 | 0.825 | 0.908 | 0.990 | 1.219 | 0.661 | 0.798 | 0.887 | 0.981 | 1.217 |
| 1000 | 0.782 | 0.889 | 0.951 | 1.016 | 1.189 | 0.772 | 0.873 | 0.934 | 1.012 | 1.172 |
| $N_{MN}$ (569) | 0.706 | 0.827 | 0.914 | 0.997 | 1.198 | 0.688 | 0.811 | 0.891 | 0.991 | 1.271 |
| $N_{DL}$ (628) | 0.735 | 0.846 | 0.926 | 1.010 | 1.184 | 0.698 | 0.834 | 0.908 | 1.002 | 1.253 |
| Scenario 3 (all different) | | | | | | | | | | |
| | $S_{MN,2,1}$ | | | | | $S_{MN,3,1}$ | | | | |
| N | 2.5th | 25th | 50th | 75th | 97.5th | 2.5th | 25th | 50th | 75th | 97.5th |
| 250 | 0.567 | 0.708 | 0.801 | 0.915 | 1.228 | 0.525 | 0.671 | 0.772 | 0.898 | 1.231 |
| 500 | 0.681 | 0.815 | 0.899 | 0.993 | 1.211 | 0.677 | 0.802 | 0.887 | 0.980 | 1.220 |
| 1000 | 0.784 | 0.880 | 0.939 | 1.015 | 1.196 | 0.771 | 0.872 | 0.941 | 1.009 | 1.175 |
| $N_{MN}$ (566) | 0.705 | 0.832 | 0.921 | 1.003 | 1.203 | 0.696 | 0.814 | 0.890 | 0.988 | 1.187 |
| $N_{DL}$ (582) | 0.714 | 0.838 | 0.911 | 1.007 | 1.253 | 0.684 | 0.817 | 0.896 | 0.985 | 1.214 |
| Scenario 4 (one rare category) | | | | | | | | | | |
| | $S_{MN,2,1}$ | | | | | $S_{MN,3,1}$ | | | | |
| N | 2.5th | 25th | 50th | 75th | 97.5th | 2.5th | 25th | 50th | 75th | 97.5th |
| 250 | 0.586 | 0.720 | 0.811 | 0.921 | 1.254 | 0.423 | 0.586 | 0.699 | 0.822 | 1.270 |
| 500 | 0.696 | 0.824 | 0.899 | 0.991 | 1.214 | 0.592 | 0.729 | 0.830 | 0.949 | 1.243 |
| 1000 | 0.789 | 0.889 | 0.952 | 1.017 | 1.165 | 0.713 | 0.837 | 0.916 | 1.005 | 1.227 |
| $N_{MN}$ (648) | 0.735 | 0.862 | 0.933 | 1.015 | 1.218 | 0.646 | 0.777 | 0.865 | 0.963 | 1.260 |
| $N_{DL}$ (901) | 0.780 | 0.880 | 0.936 | 1.011 | 1.172 | 0.689 | 0.812 | 0.888 | 0.984 | 1.220 |
| Scenario 5 (one very rare category) | | | | | | | | | | |
| | $S_{MN,2,1}$ | | | | | $S_{MN,3,1}$ | | | | |
| N | 2.5th | 25th | 50th | 75th | 97.5th | 2.5th | 25th | 50th | 75th | 97.5th |
| 250 | 0.569 | 0.708 | 0.800 | 0.917 | 1.210 | 0.260 | 0.478 | 0.594 | 0.737 | 1.118 |
| 500 | 0.681 | 0.824 | 0.909 | 0.996 | 1.194 | 0.501 | 0.660 | 0.772 | 0.897 | 1.290 |
| 1000 | 0.782 | 0.889 | 0.948 | 1.013 | 1.173 | 0.656 | 0.794 | 0.878 | 0.986 | 1.259 |
| $N_{MN}$ (706) | 0.751 | 0.856 | 0.930 | 1.019 | 1.203 | 0.586 | 0.744 | 0.846 | 0.962 | 1.291 |
| $N_{DL}$ (1458) | 0.827 | 0.915 | 0.967 | 1.019 | 1.137 | 0.717 | 0.842 | 0.913 | 1.008 | 1.214 |
| Scenario 6 (two very rare categories) | | | | | | | | | | |
| | $S_{MN,2,1}$ | | | | | $S_{MN,3,1}$ | | | | |
| N | 2.5th | 25th | 50th | 75th | 97.5th | 2.5th | 25th | 50th | 75th | 97.5th |
| 250 | 0.201 | 0.454 | 0.580 | 0.732 | 1.199 | 0.273 | 0.476 | 0.585 | 0.708 | 1.109 |
| 500 | 0.477 | 0.652 | 0.764 | 0.901 | 1.254 | 0.480 | 0.654 | 0.744 | 0.876 | 1.237 |
| 1000 | 0.636 | 0.768 | 0.861 | 0.982 | 1.282 | 0.635 | 0.775 | 0.857 | 0.960 | 1.313 |
| $N_{MN}$ (1558) | 0.719 | 0.835 | 0.924 | 1.024 | 1.277 | 0.694 | 0.824 | 0.902 | 0.995 | 1.233 |
| $N_{DL}$ (1616) | 0.725 | 0.839 | 0.927 | 1.020 | 1.285 | 0.694 | 0.834 | 0.911 | 0.998 | 1.219 |

$S_{MN,k,r}$, multinomial sub-model specific shrinkage factors.

## 11. Derivation of formula for $\max\left(R^2_{CS_app}\right)$ for multinomial models

We provide this proof as we are not aware of seeing this results in the literature.

Starting from equation (25):

$$lnL_{null} = \sum_{k=1}^{K} E_k \ln\left(\frac{E_k}{n}\right),$$

And equation (16):

$$\max\left(R^2_{CS_app}\right) = 1 - \exp\left(\frac{2\ ln\ L_{null}}{n}\right),$$

We get that:

$$\begin{aligned}
\max\left(R^2_{CS_app}\right) &= 1 - \exp\left(\frac{2 * \sum_{k=1}^{K} E_k \ln\left(\frac{E_k}{n}\right)}{n}\right) \\
&= 1 - \exp\left[\sum_{k=1}^{K} \frac{2 * E_k \ln\left(\frac{E_k}{n}\right)}{n}\right] \\
&= 1 - \prod_{k=1}^{K} \exp\left[\frac{2 * E_k \ln\left(\frac{E_k}{n}\right)}{n}\right] \\
&= 1 - \prod_{k=1}^{K} \exp\left[\ln\left(\left(\frac{E_k}{n}\right)^{\frac{2*E_k}{n}}\right)\right] \\
&= 1 - \prod_{k=1}^{K} \left(\frac{E_k}{n}\right)^{\frac{2*E_k}{n}} \\
&= 1 - \left(\prod_{k=1}^{K} \left(\frac{E_k}{n}\right)^{\frac{E_k}{n}}\right)^2 \\
&= 1 - \left(\prod_{k=1}^{K} (p_k)^{p_k}\right)^2
\end{aligned}$$

where $p_k = E_k/n$ is the is the observed frequency of category $k$, as defined in section 3.4 of the main paper.

Now, for binary logistic regression $\max(R^2_{CS_app})$ has a maximum of 0.75, when the observed outcome frequency is = 0.5.

For multinomial logistic regression, $\max(R^2_{CS_app})$ occurs when all the observed outcome frequencies are equal, i.e. $p_k = 1/K : k = 1, \ldots, K$. This results in a maximum (across all possible outcome frequencies) of:

$$= 1 - \left(\prod_{k=1}^{K} \left(\frac{1}{K}\right)^{\frac{1}{K}}\right)^2$$

$$= 1 - \left(\left(\frac{1}{K}\right)^{\frac{K}{K}}\right)^2$$

$$= 1 - \left(\frac{1}{K}\right)^2$$

$$= \frac{K^2}{K^2} - \frac{1}{K^2}$$

$$= \frac{K^2 - 1}{K^2}$$

## 12. Implications of estimating $\boldsymbol{R^2_{CS_adj}}$ through $\boldsymbol{R^2_{\text{Nagelkerke}}} = \mathbf{0.15}$ on criterion (ii)

Criterion (ii) holds if equation (22) holds. Let $R^2_{\text{Nagelkerke}}$ represent the optimism adjusted $R^2_{\text{Nagelkerke}}$ then equation (22) simplifies to:

$$S_{VH_MN} \geq \frac{R^2_{\text{Nagelkerke}} * \max\left(R^2_{CS_app}\right)}{R^2_{\text{Nagelkerke}} * \max\left(R^2_{CS_app}\right) + \delta * \max\left(R^2_{CS_app}\right)}$$

$$S_{VH_MN} \geq \frac{R^2_{\text{Nagelkerke}}}{R^2_{\text{Nagelkerke}} + \delta}$$

If we assume $R^2_{\text{Nagelkerke}} = 0.15$ when defining $R^2_{CS_adj}$, the right hand side of this equation $= 0.75$ and this criterion will always hold when $S_{VH_MN}$ is targeted at a threshold of 0.9. This is unlikely to pose an issue as $R^2_{\text{Nagelkerke}} = 0.15$ is a very safe assumption leading to high sample sizes, but it does make criterion (ii) obsolete. This is important for multinomial logistic regression as we have not been able to make any recommendations on how to estimate $R^2_{CS_adj}$, except through the assumption $R^2_{\text{Nagelkerke}} = 0.15$. It provides extra weight to the recommendation that researchers report $R^2_{CS}$ estimates of multinomial models when fitting them.